\documentclass[fleqn,10pt]{wlscirep}
\usepackage[T1]{fontenc}
\usepackage{bm}
\usepackage{comment}
\usepackage{upgreek}
\usepackage{soul}
\usepackage{enumitem}
\usepackage{graphicx}
\usepackage[font=footnotesize,labelfont=bf]{caption}
\usepackage{chngcntr}
\usepackage{hyperref}
\usepackage{etoc}
\usepackage{float}

\title{Few-Shot Neuromorphic Vision in a Nonlinear Photonic Network Laser}

\author[1,$\dagger$]{Wai Kit Ng}
\author[1,2,$\dagger$]{Jakub Dranczewski}
\author[1,2,$\dagger$]{Anna Fischer}
\author[1,3]{T. V. Raziman}
\author[1]{Dhruv Saxena}
\author[1]{Tobias Farchy}
\author[1,4]{Kilian Stenning}
\author[1]{Jonathan Peters}
\author[2]{Heinz Schmid}
\author[1,4]{Will R. Branford}
\author[3]{Mauricio Barahona}
\author[5,6]{Kirsten Moselund}
\author[1,*]{Riccardo Sapienza}
\author[1,4,7,*]{Jack C. Gartside}

\affil[$\dagger$]{These co-first authors contributed equally}
\affil[1]{Blackett Laboratory, Department of Physics, Imperial College London, London, United Kingdom}
\affil[2]{IBM Research Europe -- Zürich, Säumerstrasse 4, Rüschlikon, 8803, Switzerland}
\affil[3]{Department of Mathematics, Imperial College London, London, United Kingdom}
\affil[4]{London Centre for Nanotechnology, Imperial College London, London, United Kingdom}
\affil[5]{Laboratory of Nano and Quantum Technologies, Paul Scherrer Institut, Switzerland}
\affil[6]{INPhO, Faculty of Engineering, Ecole Polytechnique Fédérale de Lausanne, Switzerland}
\affil[7]{Institute for Materials Research, Tohoku University, Japan}
\affil[*]{Corresponding author e-mails: j.carter-gartside13@imperial.ac.uk, r.sapienza@imperial.ac.uk}

\begin{abstract} 

With the growing prevalence of AI, demand increases for hardware that mimics the brain’s ability to extract structure from limited data. In the retina, ganglion cells detect features from sparse inputs via lateral inhibition, where neurons antagonistically suppress activity of neighbouring cells. Biological neurons exhibit diverse heterogeneous nonlinear responses, linked to robust learning and strong performance in low-data regimes.

Here, we introduce a retinally-inspired photonic computing system where spatially-competing lasing modes in a random network laser act as heterogeneous, inhibitively-coupled neurons - enabling feature detection, few-shot classification, and segmentation. 

This silicon-compatible scheme harnesses heterogeneous excitatory and inhibitory nonlinear physical dynamics which give rise to emergent photonic computing behaviour, including parallel feature detection and strong performance when training data is scarce. We report 98.05\% and 87.85\% accuracy on MNIST and Fashion-MNIST, and 90.12\% on BreaKHis cancer diagnosis - outperforming software CNNs including EfficientNetV2 and the vision transformer ViT in few-shot and class-imbalanced regimes with training sets of up to several hundred images. We demonstrate combined segmentation and classification on the HAM10k skin lesion dataset, achieving DICE and Jaccard scores of 84.49\% and 74.80\%. These results demonstrate the potential of random lasing networks as nonlinear photonic learning systems, and highlight the ability of heterogeneous nonlinear dynamics to support strong learning in challenging low-data scenarios.
\end{abstract}

\begin{document}

\flushbottom
\maketitle
\thispagestyle{empty}

\section*{Introduction} 

Artificial intelligence (AI) is increasingly ubiquitous across society and industry. Demand grows for dedicated AI hardware capable of strong performance in use cases where training data and energy are scarce, such as edge computing applications where limited data is collected locally and the power and cooling overheads of GPUs are unattractive. These tasks, including image classification, medical diagnosis, and segmentation, require a range of capabilities: 
\begin{itemize}
    \item Strong nonlinearity – crucial for harder tasks\cite{markovic2020physics,mcmahon2023physics,shastri2021photonics}.
    \item Extraction of a range of informative features from input data – neural networks capable of this such as convolutional neural networks (CNNs) \cite{lecun1995learning,o2015introduction} and transformers\cite{vaswani2017attention,khan2022transformers} typically outperform those without, e.g. multilayer perceptrons (MLPs) \cite{lecun1995learning,o2015introduction,li2021survey,gu2018recent}.
    \item The ability to rapidly learn from few training examples – termed the few-shot/low-data regime \cite{song2023comprehensive,tyukin2021demystification}.
\end{itemize}

Physics-based neuromorphic computing systems\cite{markovic2020physics,momeni2024training,shastri2021photonics,farmakidis2024integrated,jebali2024powering,allwood2023perspective,manneschi2024noiseawaretrainingneuromorphicdynamic, oguzProgrammingNonlinearPropagation2024} have been explored as candidates for edge-AI hardware\cite{jebali2024powering}, but integrating the above qualities to achieve software-equivalent performance on harder problems with very limited training data, such as biomedical tasks, remains challenging.
 
Biological systems offer a compelling blueprint. In the retina, ganglion cells perform spatially-sensitive nonlinear processing via lateral inhibition\cite{kerschensteiner2022feature,masland2012neuronal,thoreson2012lateral,werblin2011retinal}, where neurons compete to fire while inhibitively suppressing the activity of neighbouring cells. This mechanism enables efficient encoding of spatial structure, even under sparse or noisy input, and the ability to detect many features in parallel from multiple ganglion cells\cite{kerschensteiner2022feature,fairhall2006selectivity}. The combination of excitatory dynamics - firing in response to input, and inhibitory ones - suppressing activity of nearby coupled neurons, is crucial to the strong performance of biological learning systems, but to date physics-based neuromorphic schemes typically focus solely on engineering excitatory dynamics. Addressing this inhibition is an inviting route towards taking neuromorphic performance closer to the biological systems which inspire it. In addition, the diverse heterogeneous nonlinear responses across ensembles of biological neurons have been linked to robust learning\cite{perez2021neural}, and creating effective high-dimensional latent spaces which have been shown to give rise to strong few-shot learning in software networks\cite{tyukin2021demystification}.

Photonic computing schemes have demonstrated promising routes to high-bandwidth, low-energy performance, with the realisation of efficient optical nonlinearities identified as a key area requiring further development\cite{mcmahon2023physics, shenDeepLearningCoherent2017, savageLightCouldLower2026}. 
Recent studies have shown that access to strong nonlinearities can substantially enhance computational capability, for example through opto-electronic nonlinear materials\cite{wang2023image}, nonlinear mode mixing in multi-mode fibres\cite{oguzProgrammingNonlinearPropagation2024}, or nonlinear dynamics such as lasing mode competition in multi-mode VCSELs\cite{porteCompleteParallelAutonomous2021, skalliModelfreeFronttoendTraining2025}. 
Exploration of combining heterogeneous nonlinearities is particularly promising here, integrating multiple distinct nonlinear dynamics - which has been shown to give rise to robust learning in software models of biologically-inspired neurons\cite{perez2021neural}.
Developing physical learning systems which exhibit both strong heterogeneous nonlinearities and the capability for useful feature extraction is hence highly attractive, with experimental demonstration of schemes integrating these qualities so-far elusive.

Here, we present a retinally-inspired neuromorphic vision system exhibiting strong heterogeneous photonic nonlinearities, parallel feature detection, and strong learning performance in the few-shot/low-data regimes where training examples are scarce. Our system is implemented in a 150 $\upmu$m diameter on-chip semiconductor random network laser\cite{Saxena2024designed}, fabricated through the mature process of wafer-bonding InP on oxide-coated Si, followed by electron beam lithography and reactive ion etching\cite{dranczewskiPlasmaEtchingFabrication2023}. 

These complex random lasers host many strongly interacting lasing modes\cite{cao2019complex,sapienza2022controlling,wiersma2008physics,saxena2022sensitivity}, with nonlinear physics well-suited for use in a neuromorphic system. The large population of modes is spatially and spectrally varying with complex disordered distributions, and provides strong intrinsic heterogeneous photonic nonlinearity - including both \textit{excitatory} dynamics, where modes reach their threshold and lasing activates, and \textit{inhibitory} dynamics, where spatially-overlapping modes antagonistically compete for finite optical gain (termed `mode competition')\cite{cerjan2016controlling,saxena2022sensitivity,van2007spatial,bachelard2014adaptive}. This combination of competing processes mirrors the coupled excitatory and inhibitory dynamics of retinal ganglion cells, where cells attempting to activate in response to incoming light are laterally inhibited by neurotransmitters from neighbouring cells. In the retina, this directly gives rise to biological feature and edge detection\cite{fairhall2006selectivity,werblin2011retinal,thoreson2012lateral,masland2012neuronal,kerschensteiner2022feature}.

The networks' sensitivity to structured optical input\cite{cerjan2016controlling,bachelard2014adaptive,saxena2022sensitivity,gaio2019nanophotonic} makes them an intrinsically well-matched platform for spatial processing. We demonstrate that the competing excitatory and inhibitory nonlinear photonic dynamics give rise to parallel image feature detection, in addition to the higher-level downstream vision tasks of classification and segmentation -- with strong performance in few-shot/low-data regimes. Crucially, the system outperforms a range of software neural networks when training data is scarce, including a pre-trained vision transformer\cite{dosovitskiy2020image} (ViT\_b\_16, 86 million parameters) and a pre-trained large-scale CNN\cite{dosovitskiy2020image} (EfficientNetV2-B0, 7.9 million parameters) on hard biomedical tasks with scarce and class-imbalanced training data.
 
These results motivate the integration of heterogeneous nonlinear physical dynamics and feature extraction capabilities in physics-based learning systems as a highly-promising direction for performing challenging few-shot/low-data learning tasks, crucial to deliver the growing demand for edge-computing systems capable of performing both inference and training\cite{chen2019deep,zeng2024efficient}. Additionally, we show that random network lasers can provide such integrated capability in on-chip photonic systems. By embedding bio-inspired behaviour combining excitation, inhibition, and nonlinear neuronal heterogeneity directly into the physical dynamics, we enable neuromorphic processing that meets the demands of edge AI.

\begin{figure}[t!]
\centering
\includegraphics[width=0.88\textwidth]{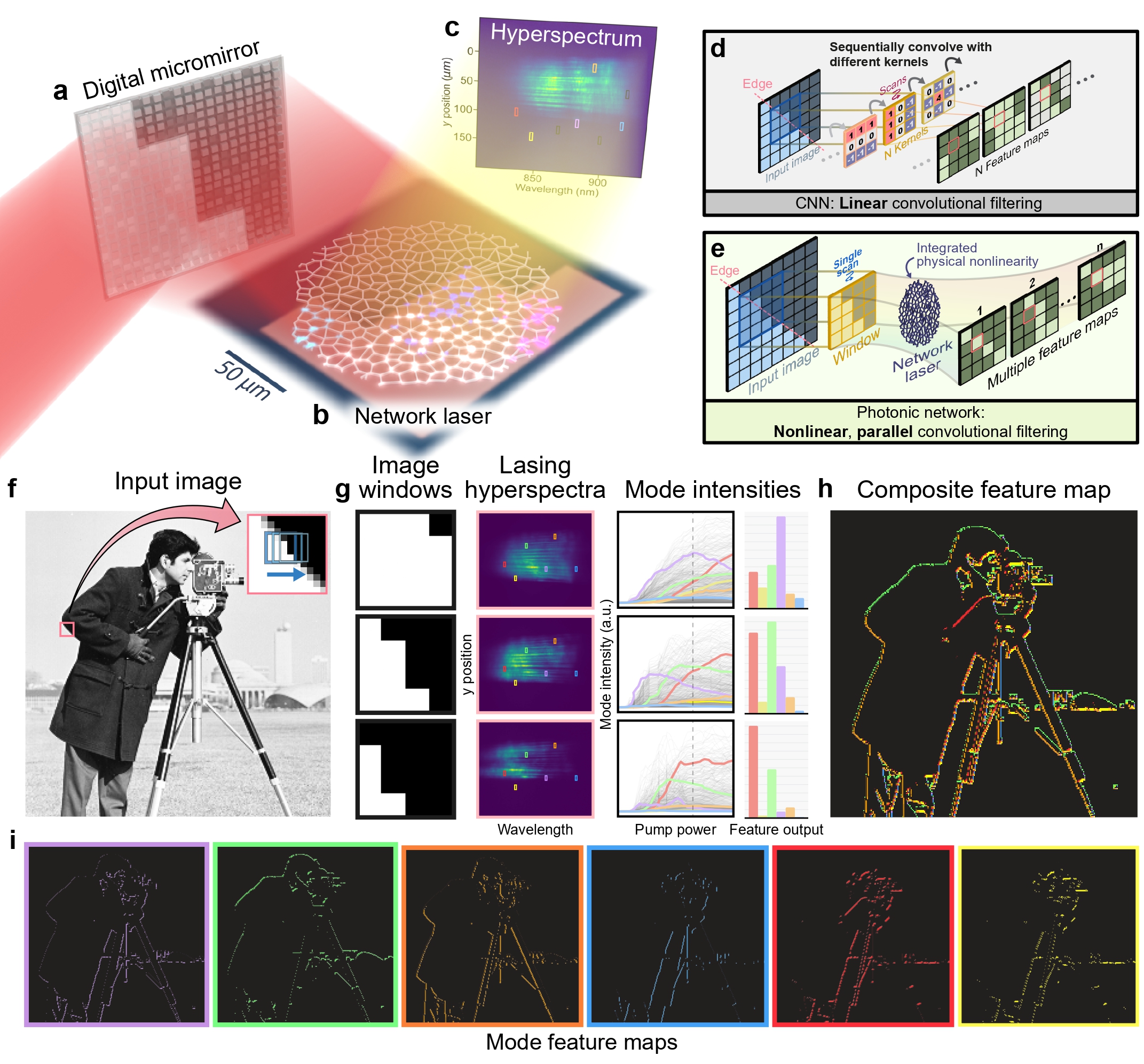}
\caption{\textbf{Working principle of neuromorphic network laser vision scheme.} \textbf{a)} Digital micromirror device spatially structures 633 nm, 200 fs pulse pump light into an arbitrary input image, which illuminates a 150 $\upmu$m InP network \textbf{b)}. The pump light triggers optical gain in illuminated network regions, exciting photons which scatter through the network to produce many lasing modes at different wavelengths determined by the scattering paths. Spatial profiles of three lasing modes are indicated via different colour network nodes. \textbf{c)} Lasing spectra emitted by the network are recorded via spatially-resolved spectrometer, giving `hyperspectral' data with spectral and spatial dimensions, at $\sim$1 $\upmu$m spatial resolution. Spectral regions which detect image features are highlighted by coloured boxes, corresponding to feature maps in h) and i). \textbf{d)} Schematic of feature detection via kernel filters in software convolutional neural networks. Linear kernel filters are raster-scanned over the input image in small `windows', performing sequential multiply-and-accumulate operations to produce feature maps. Each image feature (e.g. left or right edge) is detected by a separate kernel filter sequentially. As the process is linear, subsequent nonlinear activations must be performed on feature maps to enable vision tasks such as image classification.
\textbf{e)} Schematic of network laser feature detection. Small windows of input image are illuminated on the network in a raster-scan. Different lasing modes detect image features in parallel, with integrated physical nonlinearity. Experimentally, ten features are concurrently detected. \textbf{f)} Input `Cameraman' image, inset shows example 4$\times$4 image windows (blue boxes) and raster-scan direction. \textbf{g)} Image windows from f) with corresponding network spectral output (2nd col.), LL plots showing the heterogeneous nonlinear response of network modes in response to the different input image pump illuminations (3rd col.), and histograms showing mode intensities for six feature-detecting lasing modes (4th col.). Mode amplitudes vary in response to the presence of different image features, enabling feature detection. \textbf{h)} Composite feature map detected by the InP network. Different feature colours correspond to the lasing amplitude of the six lasing modes. \textbf{i)} Feature maps from the six lasing modes which form the composite map (h), highlighting the range of parallel features detected.
}
\label{Fig1} 
\end{figure}

\section*{Results and Discussion}

\subsection*{Working Principle}

Our physical system is a 150 $\upmu$m diameter InP network laser comprising interconnected lateral nanoscale waveguides, lithographically patterned and etched from a Si-bonded InP layer\cite{Saxena2024designed,dranczewskiPlasmaEtchingFabrication2023}. The network has a random Voronoi topology with three waveguides meeting at each vertex (average waveguide length = 5 $\upmu$m). Fabrication and optical setup details are provided in the Methods and supplementary figure \ref{setup}. A pump beam, spatially-structured via digital micromirror device (DMD) (figure \ref{Fig1}a), induces optical gain in the InP network (fig. \ref{Fig1}b). Photons are generated by spontaneous emission and scatter through the waveguides, amplified by stimulated emission from the illuminated InP. Constructive interference along specific scattering paths defines optical modes, with the vast number of potential scattering paths giving rise to a large number of spatially-distributed distinct overlapping lasing modes which nonlinearly compete for gain. The modes undergo net gain or loss depending on the extent of overlap between the illumination pattern and the mode profiles. The $\sim$100 nm bandwidth of the lasing spectra is given by the gain spectrum of the network material (InP), roughly centred around the bandgap.

Pump illumination is supplied by a 633 nm, 200 fs pulse laser, with pulse energy measured at 43 nJ at the sample (under full illumination). The spatial distribution of light emitted from the network by three different modes is illustrated by the coloured network nodes in figure \ref{Fig1}b), experimentally imaged by a frequency-resolved slit-scan camera. The network lasing response is measured using a two-dimensional CCD detector with diffraction grating. This allows collection of two dimensional hyperspectra (fig. \ref{Fig1}c) with one spectral dimension, and one spatial dimension at $\sim$ 1$\upmu$m spatial resolution, and each hyperspectrum is acquired by a single camera image acquisition. The broad lasing modes in the spectral dimension originate from band-filling and rapid carrier-induced index changes caused by ultrashort pulse excitation. Coloured boxes on the hyperspectrum correspond to the location of six feature-detecting modes examined in figure \ref{Fig1} f-i) (additional details in supplementary figure \ref{microscope-image-spectra}).

Figure \ref{Fig1}d) shows a schematic of feature detection in conventional software CNNs. A kernel filter is raster-scanned across small image `windows' of an input image. The kernel values are matrix multiplied against image pixel values, and the results summed to a single number which becomes the pixel value of an output feature map, centered at the image window location. This process is repeated across the input image to build feature maps, with different kernel filters sensitive to different image features e.g. left or right edges. To detect multiple image features, the process is repeated sequentially using different kernels - consuming time and energy. This process is entirely linear, with subsequent nonlinear activations performed at time and energy cost in order to use the feature maps for higher-level AI tasks such as image classification or segmentation. This linear matrix multiplication approach is what has so-far been employed in physical convolutional feature detection schemes\cite{feldmann2021parallel,xu202111,yao2020fully,zheng2024multichannel}.

Figure \ref{Fig1}e) shows a schematic of the network laser feature detection. Image windows are illuminated on the network in a raster-scan fashion similar to CNNs, with the hyperspectral channels of the network’s lasing emission acting as the output. While input parallelisation through input wavelength multiplexing has been achieved in photonic schemes\cite{feldmann2021parallel}, we achieve feature detection parallelisation via the output spectral channels, where different modes at different wavelengths and spatial positions provide sensitivity to different image features. One raster-scan produces multiple feature maps, with integrated physical nonlinearity, ten experimentally and forty in simulation. Our network occupies a compact 150 $\upmu$m footprint on chip, alongside other off-chip optical components (e.g., laser source, DMD, and camera, see supplementary information).

Figure \ref{Fig1}f-i) show experimental parallel feature detection results for a `Cameraman' test image (f). Example image windows (4$\times4$ pixels, inset of f) are shown in (g) alongside their respective lasing hyperspectra (2nd col.),
light-in versus light-out (LL) plots (3rd col., network lasing mode intensity vs pump intensity), and a histogram of mode intensities (4th col.) for six feature-detecting modes, taken at the pump power indicated by dashed vertical lines in the LL plots.

Mode lasing intensity is increased or inhibited by the presence of different image features, due to a combination of correlation between the input image shape and spatial mode profile, and nonlinear mode competition between spatially overlapping modes. This above-threshold mode competition is intrinsic to multimode lasers. As overlapping modes compete for gain, they inhibit each other by spatial and spectral `hole burning' effects\cite{sapienzaDeterminingRandomLasing2019, tureci2008science}. These effects can be modelled using steady-state ab-initio laser theory (SALT)\cite{ge2010steady}, which we have previously used to build a graph-based theoretical framework of network lasers - termed ‘netSALT’\cite{saxena2022sensitivity}. The large variety of spatially and spectrally varying modes in our random network laser gives rise to strong mode competition, and high sensitivity to pump patterns (e.g. input images)\cite{saxena2022sensitivity, Saxena2024designed}. The resulting increase or inhibition of lasing modes with increasing pump power is clearly seen in the LL-plots in figure\ref{Fig1}g), which show strongly heterogeneous nonlinear responses for each mode, which vary substantially depending on the input image.

This combination of excitatory and inhibitory nonlinear responses to input images enables feature maps to be obtained by plotting the lasing amplitude of different modes above the uniform illumination level at different positions in the raster-scan. The mode intensities plotted in the histogram in (g) are used to generate the pixel amplitudes of our feature maps. Figure \ref{Fig1}h) shows a composite feature map produced by six lasing modes, with different colours corresponding to the lasing amplitude of six modes indicated in fig. \ref{Fig1}c,g-i). Figure \ref{Fig1}i) shows separate feature maps for the modes comprising the composite map. See Methods for raster-scan details. Feature maps for all ten feature-detecting modes are shown in supplementary figure \ref{allcameraman}.

\begin{figure}[t!]
\centering
\includegraphics[width=0.88\textwidth]{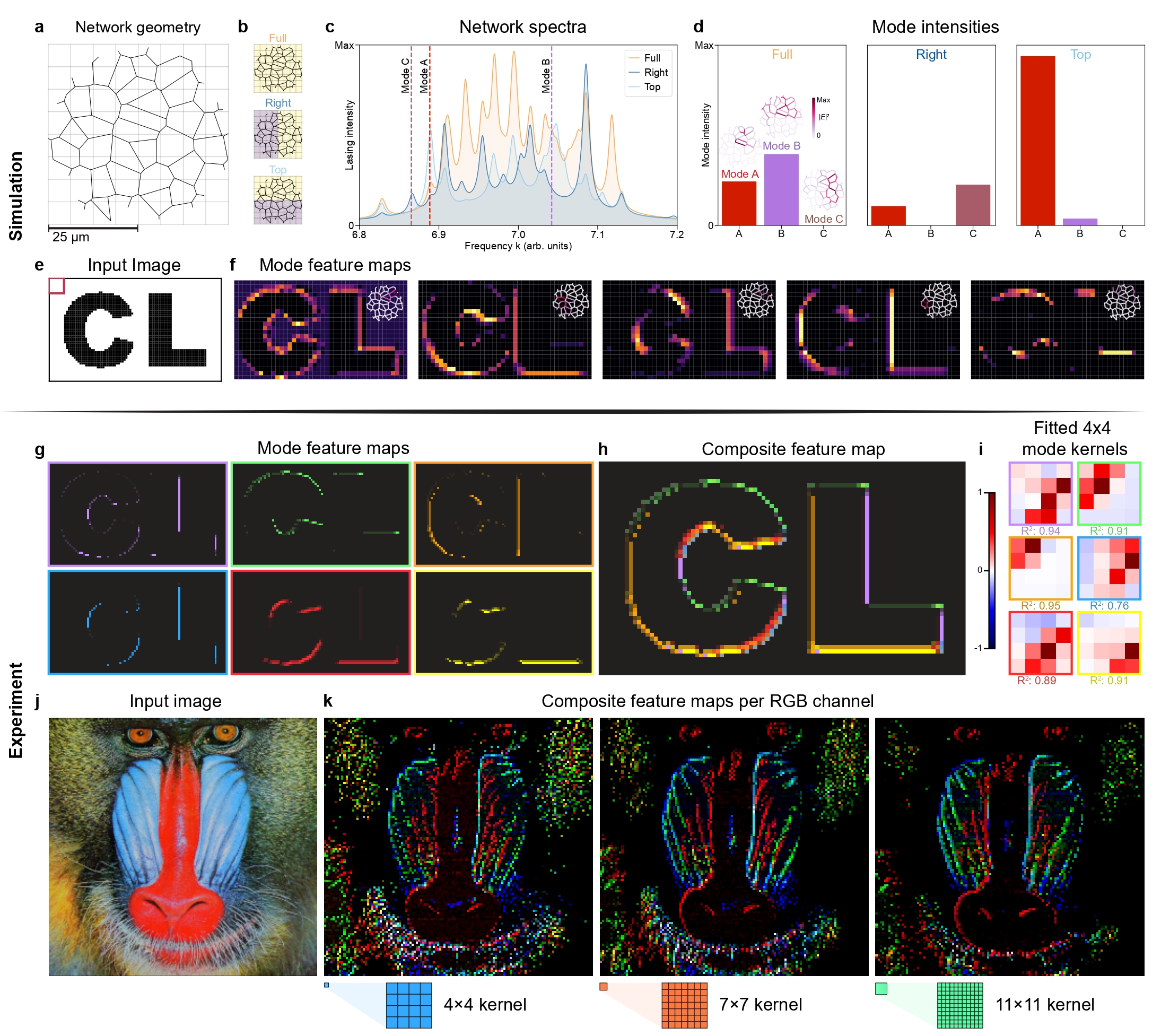}
\caption{\textbf{Parallel image feature-detection via network lasing modes.} a-f) show simulation results, g-k) are experimental.\\ \textbf{a)} 50 $\upmu$m network topology used for netSALT simulations a-f). \textbf{b)} Three pump illumination patterns: uniform, horizontal edge (`right') and vertical edge (`top'). \textbf{c)} Simulated network lasing spectra for the three illumination patterns. Three modes A-C are indicated, corresponding to d). \textbf{d)} Simulated lasing intensities of three modes A-C, for the three pump illumination patterns. Mode competition nonlinearly suppresses specific modes below lasing threshold, and amplifies other modes dependent on input pump image. Spatial mode profiles are shown inset in left panel. \textbf{e)} Input image, `CL' for `Complex Lasing'. The input image is fractured into 8$\times$8 pixel windows and illuminated onto the network sequentially in a raster-scan fashion, similar to kernel scanning in a convolutional filter process. \textbf{f)} Simulated feature maps for five modes. Pixel brightness in the feature maps corresponds to mode lasing intensity for that position in the window raster-scan. Different modes lase in response to different image features, spatial mode profiles shown inset. A larger set of 28 simulated feature maps and corresponding mode profiles is shown in supplementary information. \textbf{g)} Experimentally measured `CL' feature maps produced from six modes, with different lasing modes responding to specific image features. \textbf{h)} Experimental composite feature map combining modes from g). \textbf{i)} Equivalent 4$\times$4 kernels for the six lasing modes in (g,h), extracted by linear fits to experimental data. Coloured frames correspond to mode feature map colour scheme. \textbf{j,k)} Colour photograph of a mandrill. Image is separated into RGB channels and feature detection is performed at multiple kernel sizes demonstrating the resolution flexibility of our scheme. Composite RGB feature maps are shown in (k) for 4$\times$4, 7$\times$7 and 11$\times$11 kernels illustrating the range of high-to-low spatial frequency features detected as kernel size is increased. The sizes of each kernel are illustrated below the respective maps.
}
\label{Fig2sketch} 
\end{figure}

\subsection*{Parallel Feature-Detection via Network Lasing Modes}

We can use simulations of the photonic physics to examine the origin of the feature-detection. Here, we employ netSALT\cite{saxena2022sensitivity}, a numerical model which solves the nonlinear interaction of optical waves on graph networks within the steady-state ab-initio laser theory (SALT)\cite{ge2010steady} approximation. The netSALT model includes amplification and loss on graph edges and nonlinear competition between spatially-overlapping lasing modes. Figure \ref{Fig2sketch}a) shows a random Voronoi graph network used for the netSALT simulations. Network topology and average edge length (5 $\upmu$m) match the experimentally measured network shown in fig. \ref{Fig1}, with a smaller 50 $\upmu$m diameter used in simulation to reduce computational load. Due to the smaller network size (total edge length), the simulated network will have fewer modes than the experimental one. 

Three illumination patterns are explored (fig. \ref{Fig2sketch}b): uniform illumination (`full'), horizontal edge (`right') and vertical edge (`top'). The resultant simulated lasing spectra are shown in figure \ref{Fig2sketch}c), with strong variation in lasing spectra under the different illumination patterns. NetSALT predicts 239 possible lasing modes for the network considered here, with 172 reaching lasing threshold for at least one of the illumination patterns. The three spectra in fig. \ref{Fig2sketch}c) show the sum of all active modes for the different illumination patterns. We focus on three specific modes, labelled A,B,C, identified in figure \ref{Fig2sketch}c) by dashed vertical lines. Figure \ref{Fig2sketch}d) shows the lasing amplitude of the three modes under `full', `right' and `top' illumination patterns, with the spatial mode profiles inset in the left-hand panel. The modes are spatially-distributed across the network and overlap considerably. As multiple modes are active in the same network waveguides, nonlinear mode-competition occurs between the modes as they antagonistically compete for finite pump gain. Under full illumination, mode B shows the highest amplitude, while mode C does not lase as it falls below threshold. For a right-edge illumination, mode B now falls below the lasing threshold and is deactivated, allowing mode C to lase with the highest amplitude as it no longer faces strong mode-competition from mode B. For a top-edge illumination, mode A exhibits far higher amplitude and mode C now falls below the lasing threshold. 

It is important to note that modes A and C exhibit higher lasing amplitude under `top' and `right' edge illuminations, where the network receives less input pump energy than under uniform illumination. This is not possible in a linear system with non-interacting modes, and occurs here as removing pump light from some network regions causes modes residing even partially in those regions to lose the mode-competition interactions, e.g. mode B wins under full illumination and losing under edge illumination. This observed behaviour shows that nonlinear competition for gain between spatially-overlapping modes is providing feature-sensitive lasing dynamics, analogous to the lateral inhibition processes in retinal ganglion cells.

We now consider the netSALT-simulated feature-detection performance of the system. An input image of the letters CL (`Complex Lasing') is shown in fig. \ref{Fig2sketch}e). The image was fractured into 8x8 pixel windows and illuminated on the network in a raster-scan fashion. The lasing response was simulated for each window illumination, and the simulated mode lasing amplitude plotted at each position in the raster scan to generate separate feature-maps for each mode. Figure \ref{Fig2sketch}f) shows feature maps provided by five different modes, with the corresponding spatial mode profiles inset. These plots demonstrate the ability of the network to perform spectrally-multiplexed feature-detection of a range of image features, with the broad set of lasing modes effectively providing parallel convolutional filters. The spatial location of the mode profiles within the network are seen to show correspondence with the detected image features, e.g. modes detecting lower image edges are located towards the bottom of the network (2nd panel), modes detecting left edges are located towards the left of the network (4th panel), and so on. Of the 172 active modes in the simulated network, 40 were identified by inspection as exhibiting good feature-detection. A set of 28 selected feature maps and corresponding spatial mode profiles is shown in the supplementary figure \ref{Full-netSALT-modes}, alongside the full set of 172 feature maps from all modes in supplementary figure \ref{Full-172netSALT-modes}. A regular hexagonal network was simulated to compare against the random topology (supplementary figure \ref{netSALT-regular}), with far fewer detected features in the hexagonal network. This demonstrates the ability of network topology to control the device functionality, with further explorations into the correspondence between topology and performance providing fruitful ground for future work.

\begin{figure}[t!]
\centering
\includegraphics[width=0.88\textwidth]{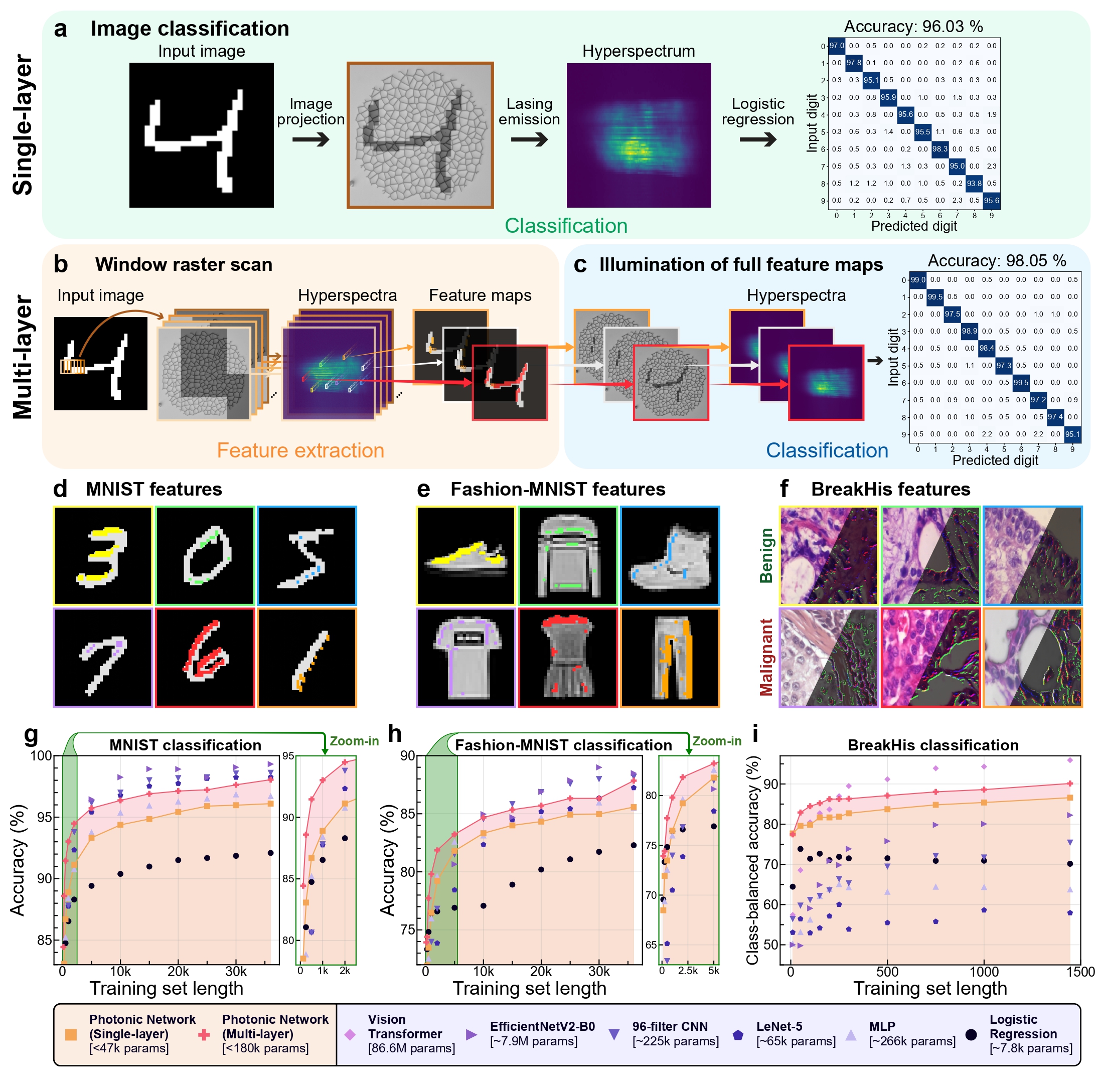}
\caption{\textbf{Network Laser Image classification.} \textbf{a)} Single-layer classification. Input images are illuminated directly onto the network, hyperspectra are measured, and a single logistic regression step is applied to the spectra to provide classification. Accuracy of 96.03\% is observed for MNIST-digits. \\ \textbf{b,c)} Multi-layer classification. An initial feature detection layer (b) is implemented as described in figure 2, generating 10 feature maps in parallel via the amplitude of ten lasing modes (indicated by coloured boxes on the hyperspectra). A second classification layer (c) is then implemented by illuminating the feature maps sequentially and measuring hyperspectra. The hyperspectra are combined, and a single logistic regression step applied to return classification predictions. Accuracy of 98.05\% is observed for MNIST-digits. \\ \textbf{d,e,f)} Three classification tasks are implemented, with example images for each task showing detected features: (d) MNIST-digits, (e) Fashion-MNIST, and (f) BreaKHis cancer diagnosis (400x magnification). \\ \textbf{g,h,i)} Classification accuracy vs. training set length, comparing single- and multi-layer photonic network accuracy against software benchmarks: logistic regression, multilayer perceptron (MLP), LeNet-5 convolutional neural network (CNN), a 96-filter CNN, EfficientNetV2-B0 CNN (3040 filters, in a transfer-learning scheme with pre-trained ImageNet weights) , and a vision transformer (ViT) with $\sim$86 million parameters. Maximum accuracies for the multi-layer photonic network of Digits: 98.05\%, Fashion: 87.85\%, and BreaKHis: 90.12\% are observed. Zoomed-in regions of the few-shot regime are shown in (g,h), where the photonic network exceeds software benchmark performance.}
\label{Fig3sketch} 
\end{figure}

We now examine experimental feature detection. Figure \ref{Fig2sketch}g) shows six experimentally measured feature maps of the `CL' image with colours corresponding to the modes shown in figure \ref{Fig1} and a composite feature map shown in figure \ref{Fig2sketch}h). Parallel feature detection of multiple image features is observed experimentally, with close correspondence to the netSALT simulated behaviour. Supplementary figures \ref{kernels_starcircle1}, \ref{kernels_starcircle2} and \ref{kernels_wave} examine the feature detection performed by ten lasing modes across a range of test images. To compare our physically-detected feature maps against linear matrix kernel filters as used in software CNNs, we extract effective kernel filters for our modes via least-squares fitting between input images and the experimentally measured mode intensity (details in supplementary information). The extracted effective filters are shown in fig. \ref{Fig2sketch}i) with further details in supplementary fig. \ref{fitted_kernels}. Notably, both positive and negative kernel weights are observed - generating bipolar weights is highly challenging in neuromorphic hardware. This is an attractive benefit of the photonic lateral inhibition processes in our system, which enable mode amplitudes to both increase and decrease in response to specific image features. The $R^2$ values below each effective kernel indicate how well the linear fitting process captures our experimental results. When performing the same kernel fitting process on arbitrary linear software kernel filters, the process performs perfectly with $R^2$ values of 1.0. The kernels extracted from our experimental data have $R^2$ values from 0.628-0.946 (see SI fig. \ref{fitted_kernels} and \ref{r2s_all_modes}), suggesting that aspects of the physical feature detection process cannot be fully captured by linear kernel filters. This is further supported by a similar linear-fit study on netSALT-simulated data shown in SI fig.~\ref{netsalt_kernel_fits}. The complexity of the fitted kernels also indicates that while these modes were chosen by inspection to produce edge-detecting feature maps, the underlying dynamics are likely to support other convolutional filters.

We extend experimental demonstration of feature detection to high-resolution colour images, showing a flexible range of kernel sizes using the popular `mandrill' test image (256 $\times$ 246 pixels, not a baboon as often thought!)\cite{levkineTestImagesCollections, usc_baboon}. Figure \ref{Fig2sketch}j) shows the input image, which is separated into RGB channels and illuminated onto the network at a range of kernel sizes to demonstrate the capability of our system to detect features with arbitrary kernel window size. This shows a flexibility of our system in handling input images and kernel windows of varying resolution.

Figure \ref{Fig2sketch}k) shows experimentally detected feature maps for 4$\times$4, 7$\times$7 and 11$\times$11 kernels, with the R/G/B coloured features detected from the corresponding colour channel (the response of all ten feature-detecting modes are summed to produce a single feature map for each colour channel here; the separate mode maps can be seen in Supplementary Figures \ref{feature_maps_rgb_k4}, \ref{feature_maps_rgb_k7}, and \ref{feature_maps_rgb_k11}). The ability to tune feature detection sensitivity to higher or lower spatial frequency is shown in our feature maps produced by different kernel sizes, the 4$\times$4 kernel picks up high-frequency details in the mandrill fur around image corners, whereas the 11$\times$11 kernel exhibits weaker response to these. Increasingly, powerful modern software CNNs such as EfficientNet\cite{tan2019efficientnet,tan2021efficientnetv2} employ a range of filter sizes, demonstrating the utility and attraction of our ability to freely select filter size. Separate lasing mode feature maps are shown for a wider range of test images in supplementary figures \ref{allcameraman}, \ref{allCL} and \ref{allStreet}.

\subsection*{Network Laser Image Classification}

Here we demonstrate the ability of the network laser to classify images. The network can be operated in a single-layer scheme, directly projecting images onto the network (fig. \ref{Fig3sketch}a), or a two-layer scheme with an initial feature-detection layer (fig. \ref{Fig3sketch}b), followed by projecting the feature-maps onto the network (fig. \ref{Fig3sketch}c). In both cases, a logistic regression step follows the final layer of hyperspectral output to generate classification predictions. 

We perform classification on three tasks: MNIST digits, Fashion-MNIST, and the challenging low-data BreaKHis breast cancer diagnosis\cite{spanhol2015dataset} (colour histopathology images, 400$\times$ magnification, 700$\times$460 resolution). Figures \ref{Fig3sketch}d,e,f) show example images for each task, with coloured lines illustrating feature maps detected by the network. For MNIST digits and Fashion, we experimentally process 40,000 training images with a test set of 4000 images and training sets of 125 to 36,000 images. For BreaKHis we process 1693 images, with a test set of 250 and training sets of 10 to 1,443 images. BreaKHis contains a roughly 2:1 ratio of malignant to benign images, with this class-imbalance increasing task difficulty. No augmentation is performed on the original datasets to synthetically increase their length.

\textbf{Single-layer classification} (fig. \ref{Fig3sketch}a) works by projecting whole images onto the network, measuring the hyperspectrum, then applying logistic regression on the training set hyperspectra to learn a fixed weight (per image class) for each channel of the hyperspectral output. This feed-forward single layer is similar in some respects to an Extreme Learning Machine, where input information is nonlinearly transformed and projected to a high dimensional state by a fixed set of complex, nonlinear nodes, and then transformed to a useful computational output via a linear readout layer\cite{HUANG2006ELM}. However, the ability of the network laser to directly produce computationally relevant outputs such as image feature maps from the intrinsic nonlinear photonic dynamics, without requiring a trained linear readout layer, means that our system goes beyond the definition of an extreme learning machine. Once the system is trained, unseen images are classified by performing a weighted sum over the hyperspectral channel amplitudes and learned weights for each class. The class with the largest weighted sum is returned as the network prediction. The single-layer scheme obtains accuracy of 96.03\% on MNIST digits, 85.18\% on Fashion MNIST, and 86.59\% on BreaKHis cancer diagnosis, with measurement throughput up to 100 Hz.

\textbf{Multi-layer classification} begins with feature detection (fig. \ref{Fig3sketch}b). Ten spectral regions performing feature detection were selected, producing ten feature maps per input image. Regions returning clear edge-detection were hand selected, with future improvement available by training the feature selection process. Three feature maps are shown schematically in figure \ref{Fig3sketch}b), with all feature maps for a range of MNIST digits and Fashion-MNIST shown in supplementary figures \ref{allDigits} and \ref{allFashion}. These feature maps are then re-illuminated on the InP network as whole images and their corresponding hyperspectra recorded, illustrated in figure \ref{Fig3sketch}c). The hyperspectral output from the second-layer feature map illuminations are combined in parallel and a single logistic regression step is applied to produce classification results. The multi-layer scheme obtains accuracy of 98.05\% on MNIST digits, 87.85\% on Fashion-MNIST, and 90.12\% on BreaKHis cancer diagnosis. Further details of the regression process are given in the Methods.

\begin{figure}[t!]
\centering
\includegraphics[width=0.89\textwidth]{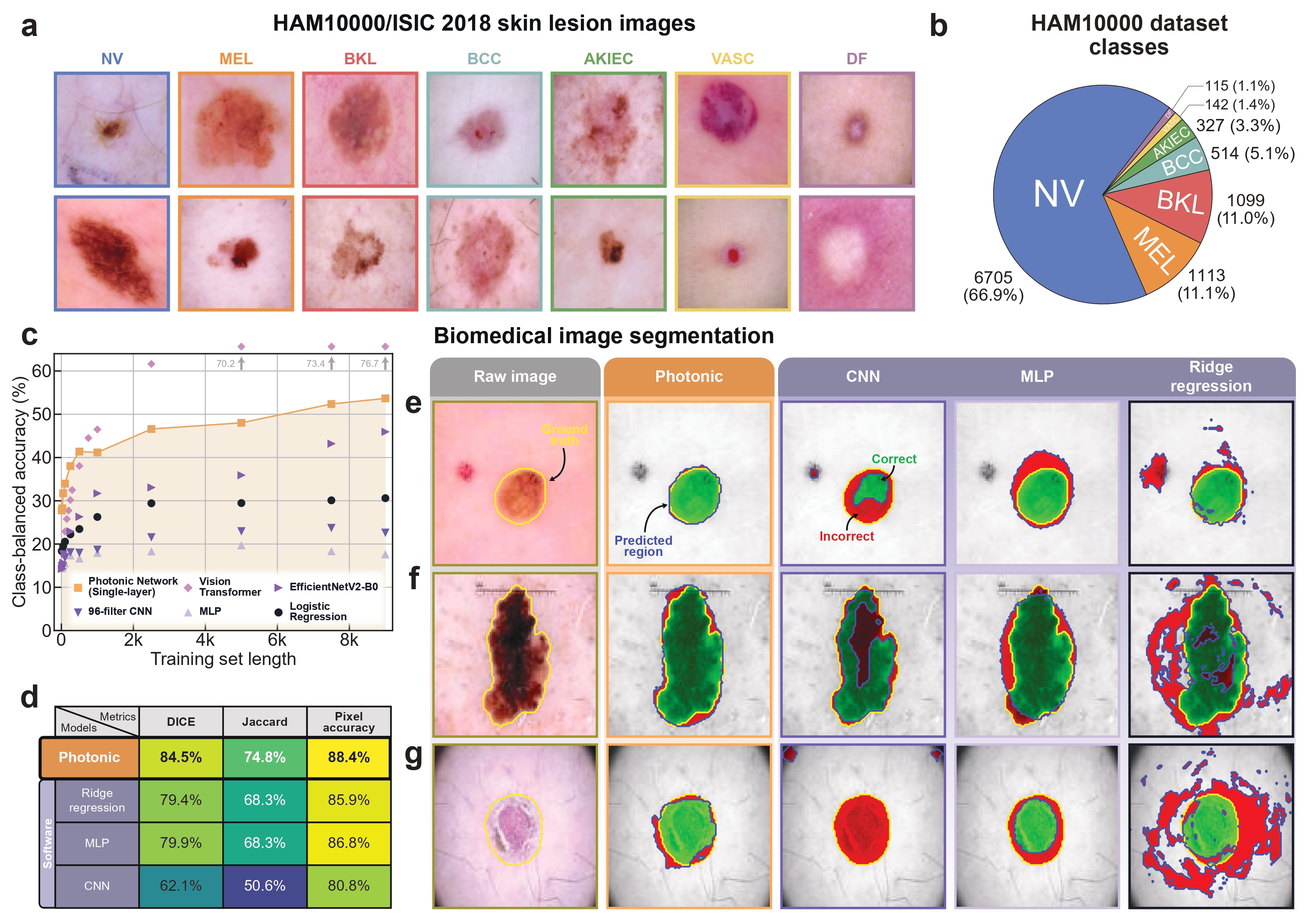}
\caption{\textbf{Biomedical Image Segmentation and Class-Imbalanced Diagnostic Classification on the HAM10,000/ISIC 2018 dataset.} \textbf{a)} Example images of the seven skin lesion classes contained in the dataset. Images are complex with different instances of the same class looking very different, hard to diagnose without expert medical training.\\ \textbf{b)} Pie-chart illustrating the heavy class-imbalance in the dataset. \\ \textbf{c)} Class-balanced accuracy vs. training set length, comparing photonic network against software benchmarks. Test set is fixed at 1015 images. Class-balanced accuracy penalises incorrect classification on rarer classes more heavily, ensuring performance is not dominated by majority classes. Photonic network scores 53.7\% max accuracy. \\ \textbf{d)} Segmentation scores: DICE, Jaccard, and pixel accuracy of photonic network and software benchmarks. 
\textbf{e-g)} Image segmentation results, comparing photonic network performance against software benchmarks for three example test set images. Ground-truth masks (yellow line) generated by a human medical expert\cite{tschandl2020human} are used for supervised training, and assessment of network performance on test images. Networks generate predicted segmentation masks (blue line) around where they think the lesion boundaries are. Green regions indicate correct segmentation where the network prediction agrees with the ground truth, red regions show combined false positive \& negative errors.}
\label{Fig4} 
\end{figure}

\textbf{Network laser performance relative to software benchmarks} is assessed in figures \ref{Fig3sketch}g,h,i). Classification accuracy vs. training set length is assessed for single- and multi-layer network laser schemes against a range of software neural networks. Five software approaches are considered for all tasks: logistic regression on the raw images (a direct control, effectively bypassing our physical system), a multi-layer perceptron (300$\times$100 neurons), a LeNet5 CNN (22 filters), a larger CNN (96 filters), and the much more powerful pre-trained `EfficientNetV2-B0'\cite{tan2019efficientnet,tan2021efficientnetv2} CNN (3040 filters, 7.9 million parameters). For the harder BreaKHis 400X task with limited data we also benchmark using a far larger and more powerful pre-trained vision transformer network\cite{dosovitskiy2020image,han2022survey,khan2022transformers} (`ViT\_b\_16') with $\sim$86 million parameters, representing a large modern software vision network. Both `EfficientNetV2-B0' and the vision transformer were trained using a transfer learning approach - initialised with pre-trained ImageNet weights, which are then updated during training. Full details of the software networks are provided in the Methods. We also benchmark against a `quasi-isomorphic' software network: an initial convolutional filter layer of 10$\times$ 4$\times$4 filters followed by 30,000 randomly weighted ReLu neurons and a final logistic regression layer, mirroring the structure of our system. Results are shown in supplementary information.

\textbf{Few-shot/low-data performance} The photonic network laser excels when training data is scarce, a regime where conventional AI can struggle. The network outperforms all considered software models in the MNIST-digits and Fashion-MNIST tasks on training sets smaller than 5,000 images (digits), 15,000 images (fashion). On the harder biomedical BreaKHis task the network outperforms the same software benchmarks, including the powerful EfficientNet CNN, across all training lengths. Only the state-of-the-art vision transformer outperforms the photonic network at longer training lengths, with the photonic network outperforming even the vision transformer up to 200 training images - a substantial result.

We score 77.7\% accuracy on BreaKHis from just \textit{10 training images} (5 examples per class), far higher than the software models. This highlights the effective learning provided by the nonlinear photonic dynamics and mode competition of the network laser. Here we compare performance when using only the raw training set. Software approaches can achieve higher scores, including on the BreaKHis task\cite{alomExplainableAIdrivenDeep2025}, but this requires costly extra steps including artificial augmentation of the training set, and extensive fine-tuning of model weights, which we do not perform in our photonic scheme.

The origin of this strong few-shot performance may lie in the system's high-dimensional, heterogeneous nonlinear responses. Recent theoretical work has shown that if a learning network is sufficiently `high dimensional' (e.g. it has a large number of output features/channels with unique nonlinear responses), then the network may learn to generalise/accurately classify from very few training examples\cite{tyukin2021demystification}. This behaviour is exhibited in our network: a broad range of unique nonlinear responses is provided by the many lasing modes, shown experimentally in figure \ref{Fig1}g). The diverse nonlinearity in the `photonic neurons' (our lasing modes) provide both high dimensionality and neuronal heterogeneity, which has been shown to promote robust learning and stronger task performance in models of biological neurons\cite{perez2021neural}.

\subsection*{Combined Biomedical Image Segmentation and Class-Imbalanced Diagnostic Classification.}

The neuromorphic capabilities of the photonic network are not restricted to feature detection and classification. Here, we demonstrate the ability of the network to spatially map regions of interest in biomedical images - drawing masks over skin lesions in colour photographic images from the HAM10000/ISIC 2018 dataset\cite{tschandl2018ham10000,tschandl2020human} (fig. \ref{Fig4}a, 10015 images of seven different classes of skin lesion). This task, termed segmentation, demonstrates that while the hyperspectral output of our network only contains information on the wavelength and one-dimensional spatial position of the lasing response, it is capable of physically encoding and processing complex two-dimensional spatial information - required for segmentation and other spatial processing. The task here is challenging: lesions are non-uniform in shape, a variety of colours and intensities, and many images contain darker non-lesion regions such as hairs, moles, and areas of uneven lighting. We combine two distinct tasks on the HAM10k dataset: class-imbalanced diagnostic classification, separating lesions into seven classes, and segmentation, drawing spatial masks to indicate lesion location. This two-fold processing has potential future impact in automated/semi-automated combined diagnosis and surgery systems. We only need to input the HAM10k images to our photonic network once to achieve both tasks, simply retraining our computationally-cheap linear output regression layer to change between diagnosis and segmentation.

\textbf{Class-imbalanced diagnosis} Here we diagnose lesion images into seven classes. This is made highly challenging by the heavy class-imbalance of the dataset - illustrated in figure \ref{Fig4}b). Machine learning typically struggles with class imbalance, tending to incorrectly classify rare classes as examples of the more common classes. Overcoming this requires costly techniques such as data augmentation, e.g. generating many distorted copies of training examples and using them to artificially extend the training set. The HAM10000 task is heavily imbalanced, with the `NV' class representing 66.9\% of the dataset, and four of the seven classes each representing 5.1\% or less. Figure \ref{Fig4}c) shows class-balanced classification accuracy against training set length (class-balanced accuracy penalises incorrect guesses on rare classes more heavily), with the photonic network scoring 53.7\% max accuracy, outperforming all software benchmarks other than the vision transformer across all training lengths, and outperforming the vision transformer at training lengths up to 500 images. This demonstrates that our photonic system exhibits strong advantages on both few-shot learning and heavily-imbalanced datasets - a hard problem with real-world benefits. 

\textbf{Biomedical segmentation} For segmentation, we use the same hyperspectral responses from the lesion images which we use for classification. Here, rather than using the diagnostic class as training/testing labels, and logistic regression as an output layer - we use the ground-truth segmentation maps as training/testing labels and ridge regression as an output layer (as the segmentation task has continuous output, rather than discrete labels). Otherwise, the procedure follows that used for classification. Figure \ref{Fig4}d) compares three segmentation metric scores across the test set: DICE (F1 score for segmentation), Jaccard (intersection over union) and pixel accuracy (correct pixels over incorrect). DICE and Jaccard provide best indication of segmentation quality. Pixel accuracy scores tend to be distorted by whether the correct ground-truth mask occupies a small or large fraction of the total image. The photonic network scores highest on all metrics, highlighting the ability of the network to implement spatial processing. 

Figures \ref{Fig4} e-g) show segmentation results for the photonic network and software benchmarks for three test images. In the top row (e), the photonic network accurately predicts the lesion location, ignoring the non-lesion mole, unlike the CNN and ridge regression software models. In the middle row (f) the photonic network outperforms software on a complex lesion shape with regions of varying intensity, and in the bottom row (g) the photonic network avoids incorrectly classifying the dark corner regions or hairs.

\section*{Conclusions}

This work highlights the benefit of introducing complex heterogeneous physical nonlinearities to photonic computing systems, particularly the combination of inhibitory and excitatory dynamics and spatially-distributed disordered mode profiles. By leveraging the ability of the disordered lasing network to provide heterogeneous nonlinear emission in both spatial and spectral domains alongside parallel feature extraction, we perform a range of vision tasks including feature detection, image classification, and segmentation, with promising performance relative to software approaches when learning with scarce training data. This bears interesting links with prior work exploring heterogeneous nonlinearity in spiking neural networks \cite{perez2021neural} and nanomagnetic systems \cite{stenningNeuromorphicOverparameterisationFewshot2024}, suggesting that engineering heterogeneous nonlinear dynamics across a broad range of photonic and non-photonic systems may provide fertile ground for the community to explore -- with potential impact on biomedical sensing, edge computing, and beyond. As the lasing modes responsible for feature detection spatially overlap, we provide parallel feature detection of multiple image features in a single network with a small on-chip footprint (with additional off-chip optical components). In contrast, existing physical convolutional hardware schemes typically detect one feature per physical repeat of the device, hence the spatial footprint, fabrication complexity and energy consumption typically scale with the number of features detected.

Strong learning in the few-shot/low-data regime with scarce training examples is a hard and open challenge in machine learning. The performance of our photonic network in this regime, including on hard biomedical tasks such as BreaKHis and HAM10k, provides valuable evidence that physics-based neuromorphic systems can handle complex, data-constrained learning tasks. These results bring physical neuromorphic schemes closer to offering viable alternatives to conventional software-based approaches. Our good performance here extends to heavily class-imbalanced datasets, which are similarly challenging for conventional software AI. One of the key proposed applications for neuromorphic computing hardware is `edge computing', where remotely deployed hardware performs computation directly at the point of sensing. While edge computing is often (though increasingly not solely\cite{chen2019deep,zeng2024efficient}) limited to inference, our system shows promise for edge deployments where both training and inference are performed, which requires adapting on-the-fly to incoming data and changing environmental conditions without access to large offline/cloud-based training data. Here, it is highly advantageous to perform well with short and imbalanced training sets, as locally collected data will typically be far from the ideal large and class-balanced training sets where software AI is most comfortable. 

A key result of this work is demonstrating the importance of including both excitatory and inhibitory interactions in neuromorphic hardware. In biological systems, both of these interactions and their interplay are crucial to providing neuronal functionality\cite{thoreson2012lateral,fairhall2006selectivity,kerschensteiner2022feature}, and existing physical neuromorphic hardware platforms often focus solely on excitatory signals. By engineering inhibitory nonlinear mode-competition between network lasing modes, we demonstrate the functional benefits of bio-inspired inhibitory dynamics.

As we develop finer control over the network dynamics, one can envision networks engineered to detect arbitrary specific features or provide reconfigurable detection via altering network dynamics through electrical control or selective illumination of network sub-regions\cite{saxena2022sensitivity} to provide effective tunable physical weights in the input space\cite{wright2022deep}. More broadly, our approach shows the benefits of designing physical computing systems with diverse physical nonlinearities, and a balance of competing excitatory and inhibitory interactions. While our system currently exhibits strongest performance on harder, lower training data tasks, we foresee that future iterations will extend this performance to larger datasets via training of the physical dynamics and optimisation of the network graph topology.

\subsection*{Author contributions}
RS, JCG, KM and KS conceived the work.\\
RS, KM, JCG, MB, KS and WRB secured funding.\\
WKN, JD, DS, TF and AF performed optical experiments.\\
JCG and KS designed the neuromorphic computing architecture.\\
JCG, JD and TVR developed the feature detection scheme.\\
RS, KM, HS, JD, AF, DS, and WKN designed and developed the network laser concept.\\
JD, KM and HS fabricated the InP network.\\
TVR performed all simulations.\\
JCG, JD, KS and TF performed machine learning analysis.\\
JD and TS performed analysis of experimental feature-detection functionality.\\
JD selected the biomedical datasets studied.\\
MB provided guidance on physical simulations using the netSALT code and guidance on implementing the machine learning architecture.\\
JP designed a benchmarking task used to develop the computing architecture.\\
AF, WKN and JD designed and produced the figures.\\
JCG, WKN, JD, and AF wrote the manuscript with contributions from all authors.

\subsection*{Acknowledgements}
This work was supported by a Royal Academy of Engineering Research Fellowship  awarded to JCG.\\
JCG was supported by the EPSRC ECR International Collaboration Grant EP/Y003276/1.\\
JCG, RS and JD were supported by the Imperial College London President's Excellence Fund for Frontier Research.\\
JCG was supported by the ERC Starting Grant MORPHON.\\
AF, JD, HM, RS and KM acknowledge support from the EU ITN EID project CORAL (GA no. 859841).\\
TVR, DS, MB and RS acknowledge support from the Engineering and Physical Sciences Research Council (EPSRC), grant number EP/T027258/1.\\
WKN acknowledges research support through a President’s PhD Scholarship from Imperial College London.\\
This work was supported by funding from the EPSRC Impact Acceleration Account awarded to JCG and RS.\\
This work was supported by the Imperial College London DT-Prime scheme awarded to JCG, RS and MB.\\
KDS was supported by The Eric and Wendy Schmidt Fellowship Program and the Engineering and Physical Sciences Research Council (Grant No. EP/W524335/1).\\
TF was supported by an EPSRC doctoral training programme PhD award.\\
JP performed the work as part of his Physics MSci project at Imperial College London.\\
The InP network laser samples were fabricated in the Binnig and Rohrer Nanotechnology Center (BRNC) at IBM Research Europe - Zurich. We thank Markus Scherrer, Daniele Caimi and the Cleanroom Operations Team of the BRNC for their help and support.\\
We thank David Mack and Stefano Vezzoli for excellent laboratory management, and Stephen Cussell and David Bowler for excellent technical workshop services at Imperial College London.\\
We thank Christine Zhao and Eunju Moon for their helpful discussions and contributions to the project.\\
Simulations were performed on the Imperial College London Research Computing Service\cite{hpc}.\\
Test image of `Cameraman' was used with CC BY-NC 4.0 creative commons licence, first appearance in Schreiber et al\cite{schreiber1978image}. Obtained from dome.mit.edu/handle/1721.3/195767.\\
Test image of `Street Scene' was used with CC0 1.0 universal creative commons licence, photographer Christian Birkholz, taken 2013. Obtained from pixabay.com/photos/calle-jose-betancort-teguise-201343/.\\ 
Test image of `CL' letters was generated by the authors.\\
Test image of `Mandrill' was taken from the Test Image Collection at \href{https://www.hlevkin.com/hlevkin/06testimages.htm}{https://www.hlevkin.com/hlevkin/06testimages.htm}, also available in the USC-SIPI image database at:
\href{https://sipi.usc.edu/database/database.php}{https://sipi.usc.edu/database/database.php}.\\

JCG, WKN, JD, TVR, DS, and KM declare their commitment to responsible scientific conduct and explicitly oppose any dual-use or otherwise harmful application of the findings presented in this work. Specifically, we strictly oppose any activities pertaining to defence, arms or military applications. 

\subsection*{Competing interests}
The authors declare the following competing interests: WKN, JD, AF, TVR, DS, KS, WB, MB, KM, RS, JCG are founders or shareholders of Rayd Technologies, a spinout company formed to commercialise the technology described in this work. A patent covering related subject matter has been published (WO2025114710A1).

\subsection*{Data and code availability}
The data used to generate the figures in this manuscript, code implementing logistic regression classification, and a set of hyperspectra for the BreaKHis 400X task are available at https://doi.org/10.6084/m9.figshare.31209973. Further hyperspectral data is available from the corresponding author on reasonable request.
The experiment automation frameworks used are open-source: https://puzzlepiece.readthedocs.io/, https://pzp-hardware.readthedocs.io/


\section*{Methods}

\subsection*{Experimental methods}

\subsection*{Fabrication}
The InP networks are fabricated on a Si chip with a 2 $\upmu$m buffer layer of SiO$_2$ according to \cite{dranczewskiPlasmaEtchingFabrication2023}. 
To obtain a defect-free InP layer on SOI, we used a Direct Wafer Bonding approach where the 127 nm thick InP layer is grown on a sacrificial III-V semiconductor wafer, with lattice-constant matched growth. This wafer is then annealed onto the SOI wafer and the sacrificial wafer removed by wet etching, resulting in a InP on SOI wafer that is cleaved into chips.
The network structures are then patterned into a spin-coated hydrogen silsesquioxane (HSQ) mask with electron beam lithography. The pattern is transferred into the InP layer using Inductively Coupled Plasma (ICP) Reactive Ion Etching (RIE)\cite{dranczewskiPlasmaEtchingFabrication2023} using a cyclic two-stage process: a CH4 and H2 plasma stage for etching, followed by O2 plasma stage for cleaning. To reduce surface state recombination losses and oxidation, the structures are passivated with phosphoric acid and coated with 3 nm of Al2O3 using Atomic Layer Deposition (ALD) \cite{ALD_Passivation_2005}.

\subsection*{Network design}

The network was designed by using a randomly-distributed point grid pattern on a plane and generating a Voronoi diagram\cite{Saxena2024designed}. A Voronoi diagram is generated by forming areas around stochastically placed points that contain all points closer to a certain point than any other point. The edges of these areas form the network design.
Each vertex of the designed network is formed by three waveguides.
The network is cropped to a circular area with 150 $\upmu$m diameter, and the density of the point patterns is varied to generate waveguide graph-edges with 350 nm width and an average length of 5 $\upmu$m.

\subsection*{Optical setup}

The edge-detection and digit classification experiments are performed in an optical microscopy setup with a spatially-structured pump (via DMD) to excite the network laser structure. A schematic of the setup is shown in the supplementary information, fig. \ref{setup}). The pump laser is a 633 nm femtosecond pulsed laser (Light Conversion PHAROS with ORPHEUS-VIS Optical Parametric Amplifier (OPA) and LYRA harmonic generator, 200 fs pulse duration, 1 kHz repetition rate) is first expanded through a telescope and shaped by reflecting on a digital micromirror device (DMD, ViALUX V-650L). The DMD consists of $\rm 1280 \times 800$ programmable mirror pixels which shape the incident uniform beam into any pixelated black-and-white image (i.e. training images). For the edge detection and MNIST classification schemes, we use the DMD only as a binary modulator, with the DMD controller thresholding the provided images at the mid-way point in brightness. For the projection of Fashion-MNIST and BreaKHis 400X images we employ a quasi-greyscale scheme, using ordered dithering with a blue noise pattern, with multiple binary DMD pixels corresponding to one greyscale data pixel. The power of the shaped pump patterns is modulated by a motorised continuously variable neutral density (ND) filter, and the exact power received on the sample is monitored by a power meter (Thorlabs, S120VC Si photodiode, 200 - 1100 nm) on the split path reflected from a glass slide inserted. The spatially-structured pump is focused on the sample through a $\rm 20\times$ objective (Olympus Plan N 20x with a 0.4 numerical aperture (NA)). The lasing emission from the network is collected through the same objective, filtered (Thorlabs FELH0750, $\rm 750~nm$ long-pass filter), and spatially and spectrally analysed using a grating spectrometer (Princeton Instruments SpectraPro HRS-300) equipped with a $\rm 600~$g/mm visible grating without an entrance slit. The signals are measured by a 2D charge-coupled device camera (CCD, Princeton Instruments Pixis 256) with a spatial resolution of $\sim 1\upmu $m.

Throughput of images was run up to 100 Hz, with the limiting factor currently the camera detection. We have run experimental tests on a similar system up to 500 Hz. With the correct camera, throughputs of 1 kHz are readily feasible without modification to the scheme. Beyond this, the next limiting factor is the rate at which the DMD can display new images. Our current DMD has 10 kHz speed, and faster DMDs are available up into the MHz, with ongoing improvements in DMD technology likely to improve this in the near future. The repetition rate of our pulsed laser goes to 100 kHz, with MHz pulsed lasers or on-chip light sources available. The eventual fundamental limit is the 10-100 ps timescale of the photonic dynamics. Beyond this, more fundamental changes to the physical system/excitation scheme will be required.

A second 80 picosecond pulse pump laser was used to confirm that we are not limited to operating with femtosecond pulses, and can operate with a range of pump wavelengths. We employed the Picophotonics CP32 532 nm MOPA pump laser, which successfully pumps our InP network to lase. Experimental data confirming this (hyperspectra, LL plots) are shown in the supplementary information.

\vspace{28pt}

\subsection*{Hyperspectral measurements}

The raw hyperspectral data recorded by the CCD has 1024 spectral channels and 256 spatial channels. For feature detection, hyperspectral regions comprising four adjacent spectral channels and eight adjacent spatial channels are used to generate each feature map, with the averaged intensity taken within the hyperspectral region. 

For the image classification layer, the hyperspectra are divided into averaged regions of sixteen adjacent spectral channels and four adjacent spatial channels, with each region passed to the logistic regression layer and assigned a single fixed linear weight during training.

The feature maps of the ``Cameraman'', ``CL'', MNIST, Fashion-MNIST, and BreakHis images (and ``street'' image in the supplementary) are generated from raster-scanning the binarised images across the kernels with a size of 4 $\times$ 4, and a stride of 1. For the ease of the data collection process and minimising redundant data collections, a set of hyperspectral data from all possible 4 $\times$ 4 image slices (a total of 65,536 excitation patterns) is collected sequentially. The feature maps of these images are then generated by picking the corresponding hyperspectra based on the slice windows. The same process applies when we generate the edge maps in the two-layer classification scheme.

The feature maps of the mandrill image (Fig. 2k) with the kernel sizes of 4 $\times$ 4, 7 $\times$ 7, and 11 $\times$ 11 are measured separately for a consistent comparison. The feature maps with RGB channels are generated by raster-scanning every image slice from the 256 $\times$ 246 pixel image (with the size of the kernel and stride of 2) on the network laser, comprising a total of $\sim$45,000 sequential acquisitions for each of the kernels.

The region-selection and averaging performed here could be implemented by using a lower resolution spectrometer, with the additional benefits of faster data acquisition and lower cost.

\subsection*{NetSALT simulations}

The lasing modes and intensities of the network are modelled using netSALT,\cite{saxena2022sensitivity} which extends steady-state ab initio laser theory (SALT)\cite{ge2010steady} for quantum graphs. 
The netSALT model was previously developed by several of the co-authors and a detailed derivation is provided in the Supplementary Information of Saxena et al.\cite{saxena2022sensitivity}.
SALT solves the following partial differential equation, derived from the Maxwell Bloch equations that describe lasing two-level atoms interacting with light, to find the lasing mode profiles $\Psi_\mu(\vec{r})$ and their frequencies $k_\mu$,
\begin{eqnarray}
\left\{\nabla^2 + \left[\epsilon_c(\vec{r}) + \frac{\gamma_a D(\vec{r})}{k_\mu - k_a + i \gamma_a})\right]\right\}\Psi_\mu(\vec{r}) = 0\,,\\
D(\vec{r}) = D_0(\vec{r}) \left[1+\sum\limits_{\nu} \Gamma_\nu \left|\Psi_\nu(\vec{r})\right|^2\right]^{-1}\,,
\end{eqnarray}
where $\epsilon_c$ is the permittivity of the medium, $k_a$ is the central frequency of the lorentzian gain profile, $\gamma_\perp$ is the gain width, $\Gamma_\nu = \gamma_a^2/[\gamma_a^2 + (k_\nu - k_a)^2]$ is the lorentzian gain at the mode frequency $k_\nu$, $D_0$ is the spatially varying pump intensity, and $D$ is the population inversion that is reduced from $D_0$ due to mode competition.
NetSALT computes the lasing from networks by solving the one-dimensional SALT equation on the edges of the network and applying continuity of fields and flux at the nodes of the network\cite{saxena2022sensitivity}.

The designed network is represented as a graph, with the edges subdivided to overlap with an 8$\times$8 grid to correspond to the kernel.
The edges of the network have a complex permittivity $\epsilon_c= (3.4 + 0.001i)^2$ when unpumped, a lorentzian gain profile centered at normalised frequency $k_a = 7$ with a width $\gamma_a = 0.35$, and are pumped up to a normalised power $D_0 = 0.5$ (For a detailed physical explanation of the parameters, see Ref.~\citeonline{saxena2022sensitivity}).
A separate simulation is performed for each 8$\times$8 binary kernel, by pumping only the edges falling under the illuminated pixels.
The passive (without pumping) modes of the network are first identified in the complex frequency plane $k$, and these are the same for all kernels.
For each kernel, the modes are tracked in the complex plane as the pump power is increased, allowing to compare between the same modes across different kernels.
As a mode reaches the real frequency plane [Im$(k_\mu) = 0$], it attains threshold and starts lasing.
Beyond this pump power, the frequency and shape of the mode do not change but mode competition between the lasing modes modifies their lasing intensities.
They further deplete the gain available to the remaining modes, increasing their thresholds.
The mode frequencies, thresholds and intensities found by netSALT incorporate the mode competition, gain depletion and nonlinearity.

As netSALT provides one mode frequencies and intensities but not the widths, the realistic spectra in Fig.~\ref{Fig2sketch}c are created by assigning a width of 0.005 to each mode and adding a broad photoluminescence background.
The mode feature maps in Fig.~\ref{Fig2sketch}f are made by selecting specific modes and plotting their intensities for the kernels corresponding to the kernel scan of Fig.~\ref{Fig2sketch}e.

\section*{Regression process for the network laser image classification task}

Hyperspectra are recorded for the ten feature-maps for each of the 40,000 input images, with initial hyperspectral resolution of 1024 spectral channels and 256 spatial channels. The top and bottom 25 spatial channels are discarded as they fall outside the InP network, leaving 206 spatial channels.

These channels are then binned using simple averaging with binning factors of 16 and 2 for the spectral and spatial channels to give 64 and 103 spectral and spatial channels respectively per feature map.

The spectral and spatial channels of all feature maps are then combined in parallel to produce a single 2D array, with each row of the array corresponding to a single input image and each column corresponding to a single `pixel' of the binned hyperspectra.

The array is separated into separate `train' and `test' sections, with final accuracy results of 98.05\% (MNIST-digits) and 87.85\% (Fashion-MNIST) reported from train and test lengths of 36,000 and 4000 images respectively. 

Feature selection is performed on the data using the SelectKBest function from Sklearn, using Chi-squared correlation between the hyperspectral data and the ground-truth image class labels in the training set. A parameter search was performed to identify 36,000 as the number of features to keep from the initial pool of 58,000 features. 

After feature selection, logistic regression is performed on the training data using sklearn's logisticRegression function. The lbfgs solver was used with a regularisation parameter of C = 6.8, identified using a parameter search. 

\section*{Regression process for the network laser image segmentation task}

The process is nominally identical to the image classification workflow, save for these details on replacing logistic regression with ridge regression, and changing the ground truth from classification labels to segmentation maps:

Ridge regression is used instead of logistic regression (using sklearn's built in ridge regression function) as we want continuous outputs here. Logistic regression returns discrete outputs, such as class-based classification predictions. Segmentation is performed by generating 2D maps of confidence/probability that a pixel is part of a lesion, hence ridge regression is used as the output layer. The confidence maps are then thresholded at 50\% confidence to produce binary maps, and regions with higher than 50\% are predicted to contain lesions. We experimented with varying thresholds and having this as an optimised hyperparameter, but the best performing threshold values were so close to 50\% (plus or minus 1\% typically) that we leave the threshold fixed at 50\% to simplify the process.

The ground truth labels used for training and testing in the segmentation tasks are 2D maps produced by a human medical expert\cite{tschandl2020human}. We train fixed weights on each hyperspectral output channel which generate segmentation maps based on network hyperspectra, trained by maximising DICE accuracy against the ground truth maps. We train on 9000 images, and test on 1015.

\section*{Software machine learning comparisons for image classification}

We benchmark our system against five software approaches for the classification tasks: logistic regression, a multi-layer perceptron, a small convolutional neural network (LeNet-5), and a larger convolutional neural network model with more filters (96) and trainable parameters, and a far larger EfficientNetV2-B0 CNN (7.9 million parameters, 3040 filters) pre-trained using ImageNet weights and operated in a transfer learning scheme.

Within the training sets, we set aside a fixed 10\% validation split to monitor generalization and detect overfitting. For CNN models, weights are initialized using Xavier/Glorot initialization (biases set to zero). We train using the Adam optimizer with learning rate $10^{-3}$, batch size 128, for 20 epochs, using cross-entropy loss. To ensure reproducibility, we report the full list of trial seeds and use a fixed dataset subsampling/splitting seed; results are aggregated across 10 independent trials.

All software CNN baseline training was performed on an NVIDIA GeForce RTX 4070 GPU. To address overfitting, validation accuracy vs test accuracy as a function of epoch (mean $\pm$ standard deviation across trials) for each dataset/model pair (MNIST-LeNet-5, MNIST-96-filter CNN, FashionMNIST-LeNet-5, FashionMNIST-96-filter CNN) are monitored. The plots of validation/test accuracy over training epochs are shown in SI Fig. \ref{MNIST_training_epochs}, and shows that validation and test accuracies track well during training, reaching a plateau without diverging. Overfitting issues are not observed.

\textbf{CNN training parameters}

The model descriptions and number of trainable parameters for each model are given below, models were coded in python using the sklearn, keras and PyTorch packages.

\begin{table}[H]
\begin{minipage}{0.95\linewidth}

\centering
\label{tab_CNNTrainPara}
\begin{tabular}{l l}
\hline
\textbf{Parameter} & \textbf{Value} \\
\hline
Total samples ($N_{\mathrm{TOTAL}}$)      & 40,000 \\
Training samples ($N_{\mathrm{TRAIN}}$)   & 36,000 \\
Test samples ($N_{\mathrm{TEST}}$)        & 4,000 \\
Validation fraction ($\mathrm{VAL}_\mathrm{{FRAC}}$) & 0.10 \\
Epochs                                    & 20 \\
Batch size                                & 128 \\
Learning rate                             & $10^{-3}$ \\
Optimizer                                 & Adam \\
Number of trials ($N_{\mathrm{TRIALS}}$)  & 10 \\
Split seed                                & 12345 \\
Trial seeds                               & $1000 + i,\; i = 0,1,\dots N_{\mathrm{TRIALS}}$ \\
Initialisation                            & Xavier (PyTorch) \\
Training hardware                         & NVIDIA 4070 GPU \\
\hline
\end{tabular}
\end{minipage}
\end{table}

\noindent\textbf{Multi-layer Perceptron}

The multi-layer perceptron comprises an input layer of 784 neurons (one per 28x28 image pixel), an initial hidden layer of 300 neurons with ReLU nonlinear activation, a second hidden layer of 100 neurons with ReLU nonlinear activation, and then a final softmax layer.

\begin{itemize}
    \item \textbf{Input Layer:}
    \setlist{nolistsep}
    \begin{itemize}[noitemsep]
        \item Neurons: 784 (one per 28x28 image pixel)
    \end{itemize}
    
    \item \textbf{Hidden Layer 1:}
    \setlist{nolistsep}
    \begin{itemize}[noitemsep]
        \item Neurons: 300
        \item Activation: ReLU (Rectified Linear Unit)
    \end{itemize}
    
    \item \textbf{Hidden Layer 2:}
    \setlist{nolistsep}
    \begin{itemize}[noitemsep]
        \item Neurons: 100
        \item Activation: ReLU
    \end{itemize}
    
    \item \textbf{Output Layer:}
    \setlist{nolistsep}
    \begin{itemize}[noitemsep]
        \item Neurons: 10
        \item Activation: Softmax
    \end{itemize}
\end{itemize}

\textbf{Number of Trainable Parameters:}
\begin{align*}
    &\text{Input Layer to Hidden Layer 1: } (784 + 1) \times 300 = 235,500 \text{ parameters} \\
    &\text{Hidden Layer 1 to Hidden Layer 2: } (300 + 1) \times 100 = 30,100 \text{ parameters} \\
    &\text{Hidden Layer 2 to Output Layer: } (100 + 1) \times 10 = 1,010 \text{ parameters} \\
    &\textbf{Total Trainable Parameters: } 235,500 + 30,100 + 1,010 = 266,610 \text{ parameters}
\end{align*}

\noindent\textbf{LeNet-5 CNN Structure:}\\

The LeNet-5 convolutional neural network architecture  comprises seven layers: two convolutional layers each followed by a max pooling layer, two fully connected layers then a final softmax output layer. In total, 22 convolutional filters are used.

\begin{itemize}
    \item Convolutional Layer 1:
    \setlist{nolistsep}
    \begin{itemize}[noitemsep]
        \item Filters: 6
        \item Kernel Size: $5 \times 5$
        \item Activation: ReLU
        \item Input Shape: $28 \times 28 \times 1$
    \end{itemize}
    
    \item Max Pooling Layer 1:
    \setlist{nolistsep}
    \begin{itemize}[noitemsep]
        \item Pool Size: $2 \times 2$
        \item Strides: $2 \times 2$
    \end{itemize}
    
    \item Convolutional Layer 2:
    \setlist{nolistsep}
    \begin{itemize}[noitemsep]
        \item Filters: 16
        \item Kernel Size: $5 \times 5$
        \item Activation: ReLU
    \end{itemize}
    
    \item Max Pooling Layer 2:
    \setlist{nolistsep}
    \begin{itemize}[noitemsep]
        \item Pool Size: $2 \times 2$
        \item Strides: $2 \times 2$
    \end{itemize}
    
    \item Flatten Layer
    
    \item Fully Connected Layer 1:
    \setlist{nolistsep}
    \begin{itemize}[noitemsep]
        \item Neurons: 120
        \item Activation: ReLU
    \end{itemize}
    
    \item Fully Connected Layer 2:
    \setlist{nolistsep}
    \begin{itemize}[noitemsep]
        \item Neurons: 84
        \item Activation: ReLU
    \end{itemize}
    
    \item Output Layer:
    \setlist{nolistsep}
    \begin{itemize}[noitemsep]
        \item Neurons: 10
        \item Activation: Softmax
    \end{itemize}
\end{itemize}

\textbf{Number of Trainable Parameters:}
\begin{align*}
\text{Convolutional Layer 1: } & (5\times5\times1 + 1)\times6 = 156 \\
\text{Convolutional Layer 2: } & (5\times5\times6 + 1)\times16 = 2{,}416 \\
\text{Fully Connected Layer 1: } & (5\times5\times16 + 1)\times120 = 48{,}120 \\
\text{Fully Connected Layer 2: } & (120 + 1)\times84 = 10{,}164 \\
\text{Output Layer: } & (84 + 1)\times10 = 850 \\
\textbf{Total Trainable Parameters: } & 156 + 2{,}416 + 48{,}120 + 10{,}164 + 850 = 61{,}706
\end{align*}

\noindent\textbf{Larger CNN model (96 filters)}

The larger CNN model has a more complex structure than the LeNet-5 CNN, with more kernel filters and trainable parameters. In total, 96 convolutional filters are used.

\begin{itemize}
    \item Convolutional layer 1:
    \setlist{nolistsep}
    \begin{itemize}[noitemsep]
        \item Input channels: 1
        \item Number of filters: 32
        \item Kernel size: $3 \times 3$
    \end{itemize}
    
    \item Max pooling layer 1:
    \setlist{nolistsep}
    \begin{itemize}[noitemsep]
        \item Kernel size: $2 \times 2$
    \end{itemize}
    
    \item Convolutional layer 2:
    \setlist{nolistsep}
    \begin{itemize}[noitemsep]
        \item Input feature maps: 32
        \item Number of filters: 64
        \item Kernel size: $3 \times 3$
    \end{itemize}
    
    \item Max pooling layer 2:
    \setlist{nolistsep}
    \begin{itemize}[noitemsep]
        \item Kernel size: $2 \times 2$
    \end{itemize}
    
    \item Fully connected layer 1:
    \setlist{nolistsep}
    \begin{itemize}[noitemsep]
        \item Input size: $64 \times 5 \times 5$ = 1600
        \item Output size: 128
    \end{itemize}
    
    \item Fully connected layer 2:
    \setlist{nolistsep}
    \begin{itemize}[noitemsep]
        \item Input size: 128
        \item Output size: 10
    \end{itemize}
\end{itemize}

\textbf{Number of Trainable Parameters}

The number of trainable parameters in the model is calculated as follows:
\begin{align*}
    \text{Conv1 parameters} &= (1 \times 32 \times 3 \times 3) + 32 = 320 \\
    \text{Conv2 parameters} &= (32 \times 64 \times 3 \times 3) + 64 = 18,496 \\
    \text{FC1 parameters} &= (64 \times 5 \times 5 \times 128) + 128 = 204,928 \\
    \text{FC2 parameters} &= (128 \times 10) + 10 = 1,290 \\
    \textbf{Total Trainable Parameters} &= 320 + 18,496 + 204,928 + 1,290 = 225,034
\end{align*}

\noindent\textbf{EfficientNetV2-B0 Transfer Learning Model}

This model uses transfer learning using the EfficientNetV2-B0 architecture with an added classification head MLP. EfficientNetV2 is a relatively modern (2019\cite{tan2019efficientnet} original EfficientNet model, 2021\cite{tan2021efficientnetv2} for the EfficientNetV2 employed here) large convolutional neural network, frequently deployed in industrial and commercial learning applications and applied to real world problems. EfficientNetV2-B0 has 3040 convolutional filters, far more powerful than the other CNNs considered here. As is common in learning applications without sufficiently large training sets to fully optimise network parameters from scratch, pretrained weights learned from training on the ImageNet dataset are used to initialise the network and set up a broad range of feature extracting kernels before the network is optimised on the training set of an arbitrary desired task.

The EfficientNetV2-B0 base model is followed by a 512 $\times$ 128 neuron multilayer perceptron classification head to aid repurposing of the initialised ImageNet weights to arbitrary tasks, comprising: global average pooling, two fully connected (dense) layers with ReLU activations, and a final softmax classification layer.

\begin{itemize}
    \item \textbf{EfficientNetV2-B0 (pretrained on ImageNet) with 512 $\times$ 128 MLP head}
\begin{itemize}
    \item Stem Convolution Layer:
    \begin{itemize}[noitemsep]
        \item Filters: 32
        \item Kernel Size: $3 \times 3$
        \item Stride: 2
        \item Activation: Swish
    \end{itemize}
    
    \item Fused-MBConv Blocks:
    \begin{itemize}[noitemsep]
        \item Stage 1: 1 block, 16 filters, expansion: 1, kernel size: $3 \times 3$
        \item Stage 2: 2 blocks, 24 filters, expansion: 4, kernel size: $3 \times 3$
        \item Stage 3: 2 blocks, 32 filters, expansion: 4, kernel size: $3 \times 3$
    \end{itemize}
    
    \item MBConv Blocks:
    \begin{itemize}[noitemsep]
        \item Stage 4: 3 blocks, 48 filters, expansion: 4, kernel size: $3 \times 3$
        \item Stage 5: 5 blocks, 64 filters, expansion: 6, kernel size: $3 \times 3$
        \item Stage 6: 8 blocks, 128 filters, expansion: 6, kernel size: $3 \times 3$
        \item Stage 7: 1 block, 160 filters, expansion: 6, kernel size: $3 \times 3$
    \end{itemize}
    
    \item Head Convolution Layer:
    \begin{itemize}[noitemsep]
        \item Filters: 1280
        \item Kernel Size: $1 \times 1$
        \item Activation: Swish
    \end{itemize}

    \item Global Average Pooling:
    \begin{itemize}[noitemsep]
        \item Reduces spatial dimensions to a single 1280-dimensional feature vector
    \end{itemize}

    \item Dense Layer 1:
    \begin{itemize}[noitemsep]
        \item Neurons: 512
        \item Activation: ReLU
    \end{itemize}

    \item Dense Layer 2:
    \begin{itemize}[noitemsep]
        \item Neurons: 128
        \item Activation: ReLU
    \end{itemize}
    
\end{itemize}
\end{itemize}

\textbf{Number of Trainable Parameters:}
\begin{align*}
    \text{EfficientNetV2-B0 Base:} & \approx 7.2 \text{ million parameters} \\
    \text{Dense Layer 1:} & (1280 + 1) \times 512 = 655,872 \\
    \text{Dense Layer 2:} & (512 + 1) \times 128 = 65,664 \\
    \text{Output Layer:} & (128 + 1) \times 10 = 1,290 \\
    \textbf{Total:} & \approx 7,922,826 \text{ parameters}
\end{align*}

\noindent\textbf{Vision Transformer (ViT-B/16) Transfer Learning Model}\\

We additionally benchmark using a Vision Transformer (ViT-B/16) image classifier, pretrained with ImageNet weights in a transfer learning scheme. Please see docs.pytorch.org/vision/main/models/generated/torchvision.models.vit\_b\_16.html for full details. Vision Transformers process an image by splitting it into fixed-size patches (here 16$\times$16 on a 224$\times$224 pixel image, linearly embedding each patch, adding a learnable class token and positional embeddings, and then applying a stack of Transformer encoder blocks with multi-head self-attention and MLP-style linear weight/nonlinear neuron sublayers. The final class-token representation is mapped to class logits via a linear classification head\cite{dosovitskiy2020image}.

In the transfer-learning scheme used here, images are resized to $224\times224$ and ImageNet-normalised. Training is performed with an initial head-only warm-up (backbone frozen) followed by end-to-end fine-tuning of all model weights, including the backbone and final MLP classifier head.

\begin{itemize}
    \item \textbf{Patch embedding:}
    \setlist{nolistsep}
    \begin{itemize}[noitemsep]
        \item Patch size: $16 \times 16$ (ViT-B/16)
        \item Input size: $224 \times 224$ $\Rightarrow$ $14 \times 14 = 196$ patches
        \item Token sequence length: $196$ patch tokens $+$ 1 class token = $197$
        \item Embedding dimension ($D$): 768
    \end{itemize}

    \item \textbf{Transformer encoder:}
    \setlist{nolistsep}
    \begin{itemize}[noitemsep]
        \item Number of encoder blocks (layers): 12
        \item Attention heads: 12
        \item MLP hidden size: 3072
        \item Normalisation: LayerNorm (with residual connections around attention and MLP sublayers)
    \end{itemize}

    \item \textbf{Classification head (fine-tuning):}
    \setlist{nolistsep}
    \begin{itemize}[noitemsep]
        \item Linear layer mapping from 768 to $K$ classes (here $K=2$ for benign/malignant)
    \end{itemize}
\end{itemize}

\textbf{Number of Trainable Parameters:}
The torchvision ViT-B/16 model has $86{,}567{,}656$ parameters for the default ImageNet 1000-class head. Replacing the head with a $K$-class linear layer gives:
\begin{align*}
    \text{Original (1000-class) head: } & (768 + 1)\times 1000 = 769{,}000 \\
    \text{Transformer trunk (excluding head): } & 86{,}567{,}656 - 769{,}000 = 85{,}798{,}656 \\
    \text{New $K$-class head: } & (768 + 1)\times K \\
    \text{For } K=2: \quad & (768 + 1)\times 2 = 1{,}538 \\
    \textbf{Total Trainable Parameters (}K=2\textbf{): } & 85{,}798{,}656 + 1{,}538 = 85{,}800{,}194
\end{align*}

\label{Bibliography}
\bibliography{Bibliography.bib}

\newpage
\appendix

\setcounter{figure}{0}

\renewcommand{\thefigure}{S\arabic{figure}}

\captionsetup[figure]{labelformat=default,labelsep=period,name={Supplementary Figure}}

\newcommand{\figref}[1]{Fig.~S\ref{#1}}

\section*{Supplementary Information}

\localtableofcontents

\subsection*{Optical setup schematic}
\addcontentsline{toc}{subsection}{Optical setup schematic}

\begin{figure}[htbp]
\centering
\includegraphics[width=0.75\textwidth]{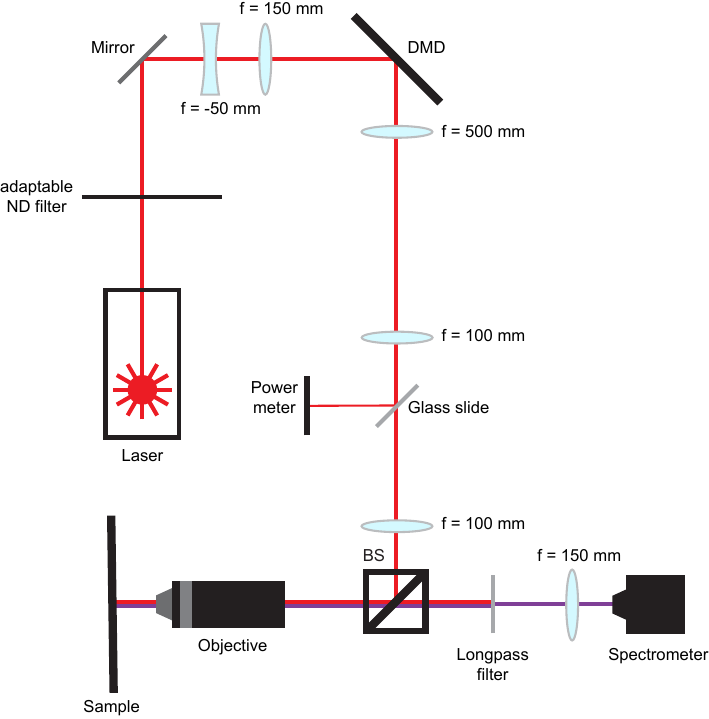}
\caption{Experimental optical setup.}
\label{setup} 
\end{figure}

Figure \ref{setup} shows a visual schematic of the experimental setup and optical path.

\subsection*{Network laser image and spectral location of feature-detecting modes}
\addcontentsline{toc}{subsection}{Network laser image and spectral location of feature-detecting modes}

\begin{figure}[htbp]
\centering
\includegraphics[width=0.9\textwidth]{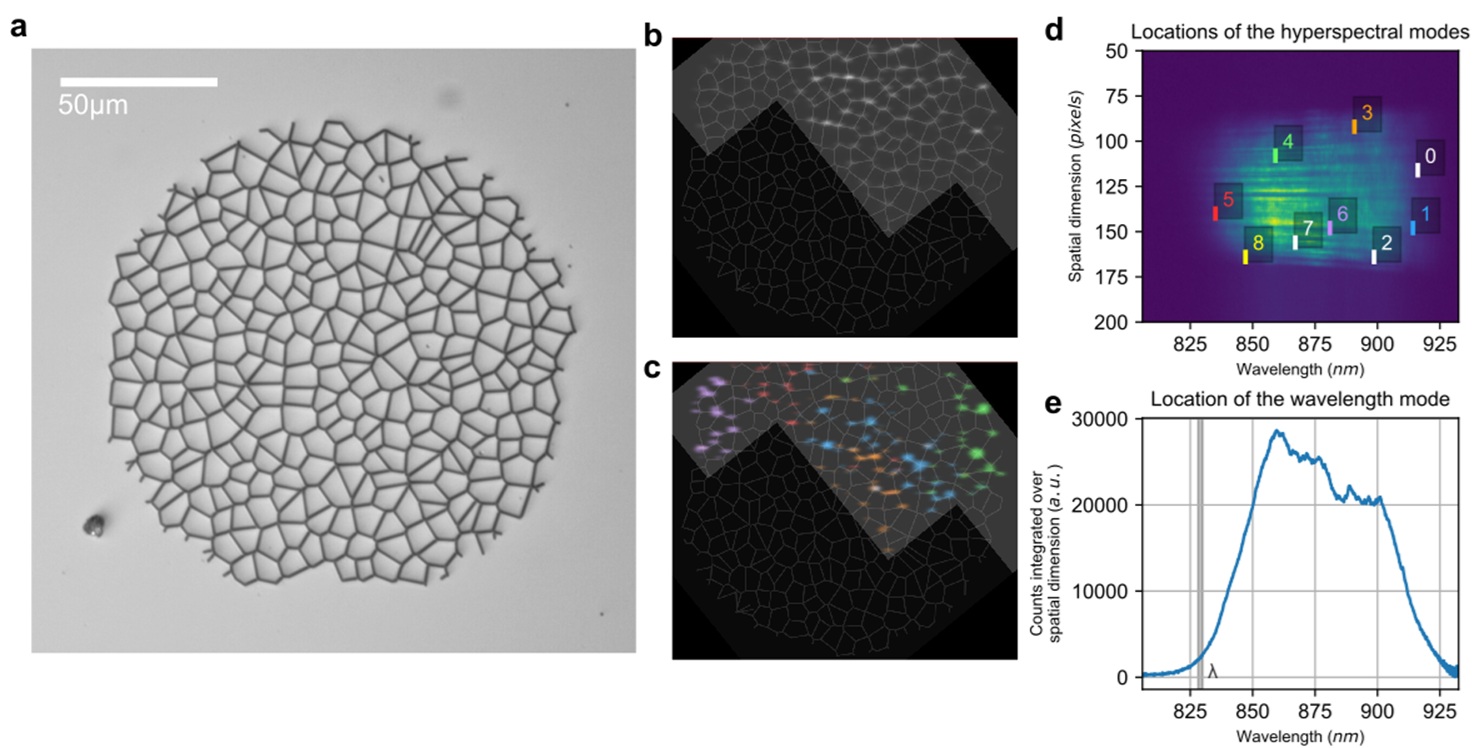}
\caption{\textbf{a)} Optical microscope image of the InP network laser used in this study.\\
\textbf{b)} Optical image of light emitted by network under diagonal-edge illumination (dark regions of the network receive no pump illumination.\\
\textbf{c)} Optical image from b) with emitted light by the network colour-coded corresponding to the wavelength of the emitted light, with the colours corresponding to the labelled hyperspectral regions in d). Different regions of the network are seen to emit light at different wavelengths.\\
\textbf{d)} Network hyperspectral output recorded for the pump illumination pattern shown in b,c). Nine hyperspectral regions are labelled, colour-coded in correspondence with the detected feature maps used throughout this study indicating the wavelength and spatial position of the spectral regions used for feature detection. Notably, feature detecting regions are largely positioned around the edges of the main central region of hyperspectral intensity.\\
\textbf{e)} Position of the tenth feature-detecting spectral region, which is formed by averaging over all spatial channels and taking the amplitude within the indicated grey region.}
\label{microscope-image-spectra} 
\end{figure}

Figure \ref{microscope-image-spectra} shows details of the InP network laser structure, spatial localisation of lasing modes and the hyperpectral regions used for feature detection. Figure \ref{microscope-image-spectra}a) shows an optical microscope image of the network used in this study. Figures \ref{microscope-image-spectra} b) and c) show the spatial profile of the light emitted by the network in response to a spatially-structured pump illumination (dark regions of the network receive no pump illumination. Figure \ref{microscope-image-spectra} b) shows all light emitted by the networks, figure \ref{microscope-image-spectra} c) shows the spatial localisation of different wavelength lasing modes, with the colour-coding corresponding to the highlighted regions on the hyperspectra shown in figure \ref{microscope-image-spectra} d). The colour corresponds to the scheme used for the mode feature maps throughout this work, eg. the red feature maps are formed via the amplitude of the hyperspectral region highlighted in red (labelled `5'). Figure \ref{microscope-image-spectra} e) shows the spectral position of the feature-detecting mode formed by averaging over all spatial channels.


\subsection*{Experimental demonstration of network lasing using picosecond pulse pump laser}
\addcontentsline{toc}{subsection}{Experimental demonstration of network lasing using picosecond pulse pump laser}

\begin{figure}[htbp]
\centering
\includegraphics[width=0.9\textwidth]{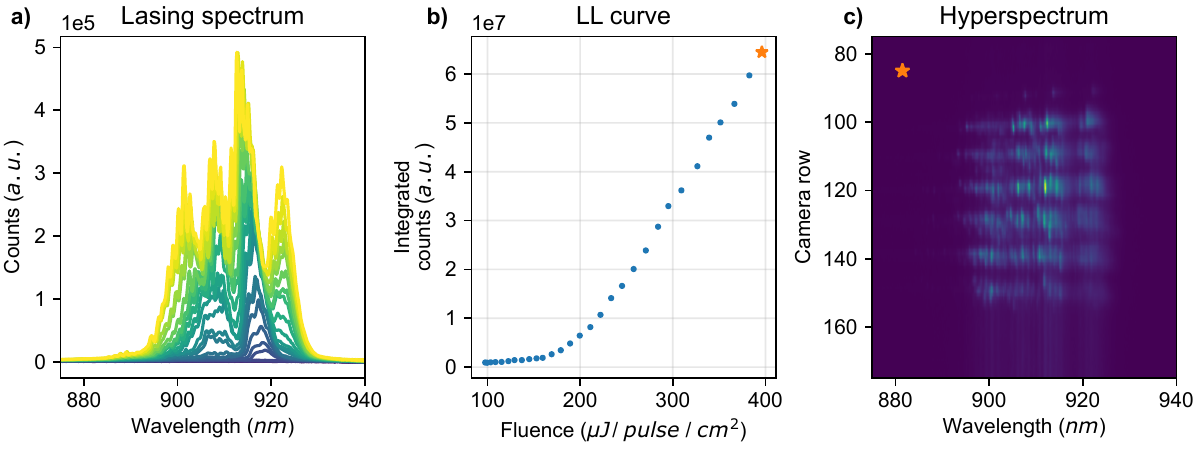}
\caption{\textbf{Experimental demonstration of InP network lasing when pumped by an 80 ps 532 nm pump laser}\\
\textbf{a)} Network lasing spectra (integrated spatially over whole network) in response to 80 ps, 532 nm pump (picophotonics brand laser) pulses of increasing intensity.\\ 
\textbf{b)} LL curve showing onset of lasing.\\
\textbf{c)} Network hyperspectrum taken at pump fluence of 400 microjoules per pulse per cm$^2$ (indicated by orange star in \textbf{b)}). Many network lasing modes are observed}
\label{picophotonic-pulse} 
\end{figure}

To confirm that we are not restricted to pumping our InP network laser using femtosecond or 633 nm pulses, here we demonstrate experimental pumping of our network using a 532 nm 80-picosecond pulse laser, supplied by Picophotonics (model CP32). Figure \ref{picophotonic-pulse} shows spatially-integrated lasing spectra, LL curve, and network hyperspectrum all confirming that the network lases in response to an 80-picosecond pump. The Picophotonics CP32 pump laser consumes 10.5 W, with a rep rate of 10 kHz.

\subsection*{NetSALT simulated feature detection - additional feature maps and spatial mode profiles}
\addcontentsline{toc}{subsection}{NetSALT simulated feature detection - additional feature maps and spatial mode profiles}

Figure \ref{Full-netSALT-modes} shows an expanded set of 28 netSALT simulated modes, spatial mode profiles and corresponding feature maps for the CL test image. A broad range of detected image features are observed.

Figure \ref{Full-172netSALT-modes} shows the full set of all 172 netSALT simulated mode feature maps. Many detected features are observed, alongside modes which only respond to very specific image regions - often lasing for only a handful of pixels of the test image eg modes 62, 157 and 109. Some modes return an output close to the input image, sometimes with enhanced lasing amplitude along a specific image feature, eg modes 2, 6, 9 and 35.

Figure \ref{Full-172netSALT-mode-intensities} shows the maximum intensities attained by the 172 simulated modes to illustrate their contribution to the output spectrum. Most of the modes show at least 1-10\% of the maximum output intensity, with the ones highlighted in Figure \ref{Full-netSALT-modes} in the 10-100\% range.

\begin{figure}[htbp]
\centering
\includegraphics[width=0.75\textwidth]{SI/netSALT_all_mode_maps_and_profiles-08-04-24.jpg}
\caption{Spectrally-multiplexed feature-maps and corresponding spatial mode profiles simulated via netSALT.
}
\label{Full-netSALT-modes} 
\end{figure}

\begin{figure}[htbp]
\centering
\includegraphics[width=1.0\textwidth]{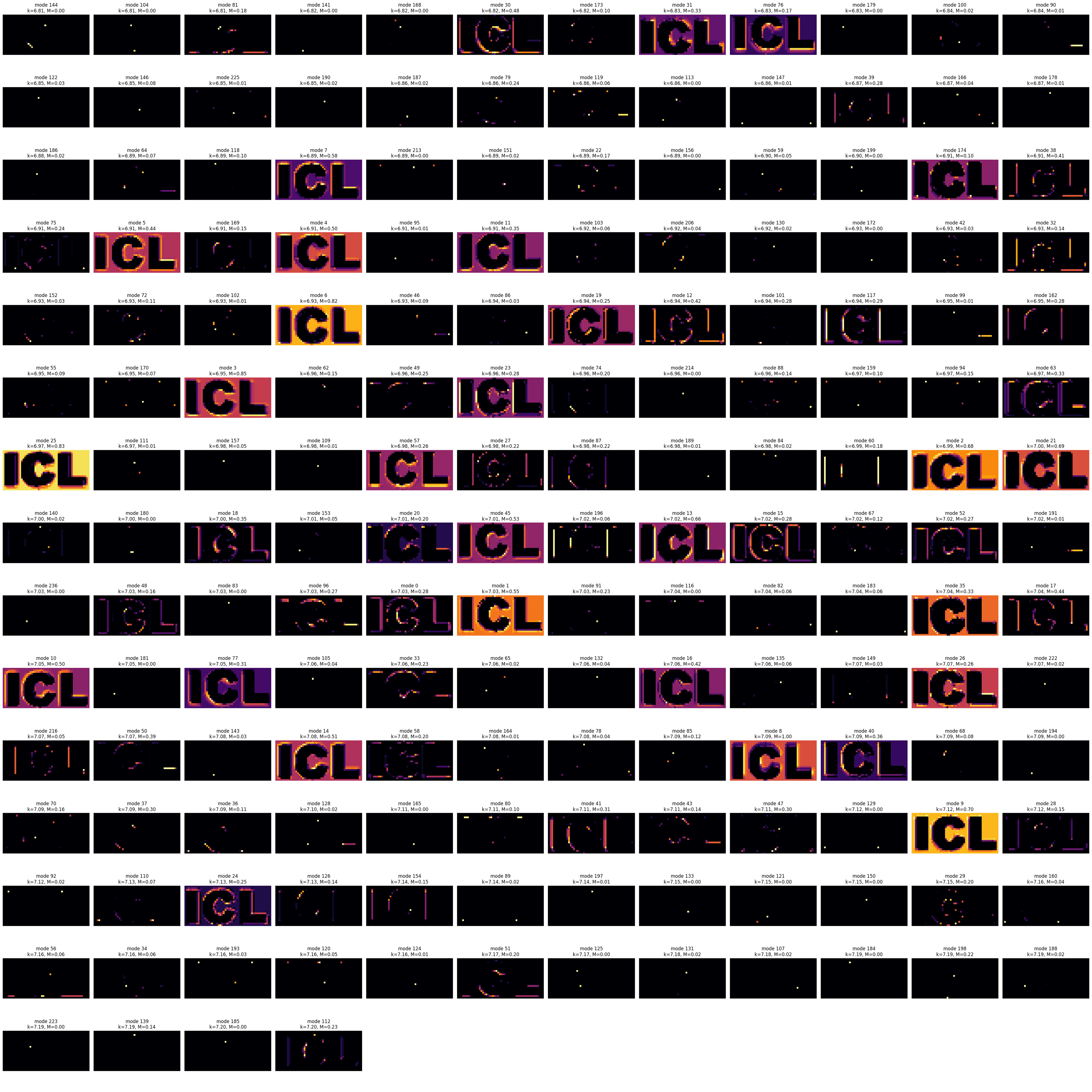}
\caption{Feature maps from all 172 netSALT simulated modes.
}
\label{Full-172netSALT-modes} 
\end{figure}

\begin{figure}[htbp]
\centering
\includegraphics[width=0.7\textwidth]{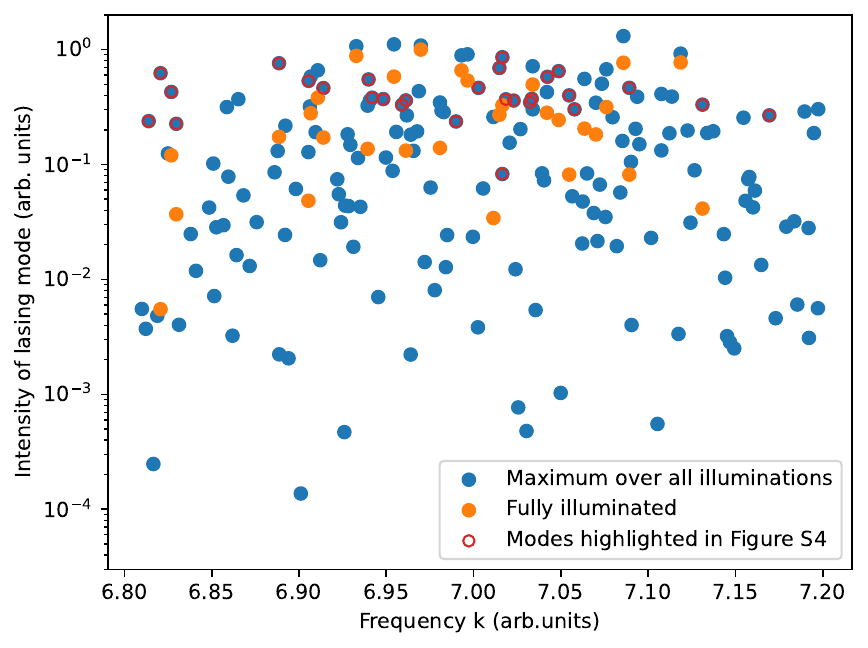}
\caption{Maximum intensity for the 172 netSALT simulated modes.
Modes with feature maps shown in Figure~\ref{Full-netSALT-modes} are highlighted.
}
\label{Full-172netSALT-mode-intensities} 
\end{figure}

\newpage
\subsection*{NetSALT simulated feature detection for a regular hexagonal network}
\addcontentsline{toc}{subsection}{NetSALT simulated feature detection for a regular hexagonal network}

\begin{figure}[htbp]
\centering
\includegraphics[width=1.0\textwidth]{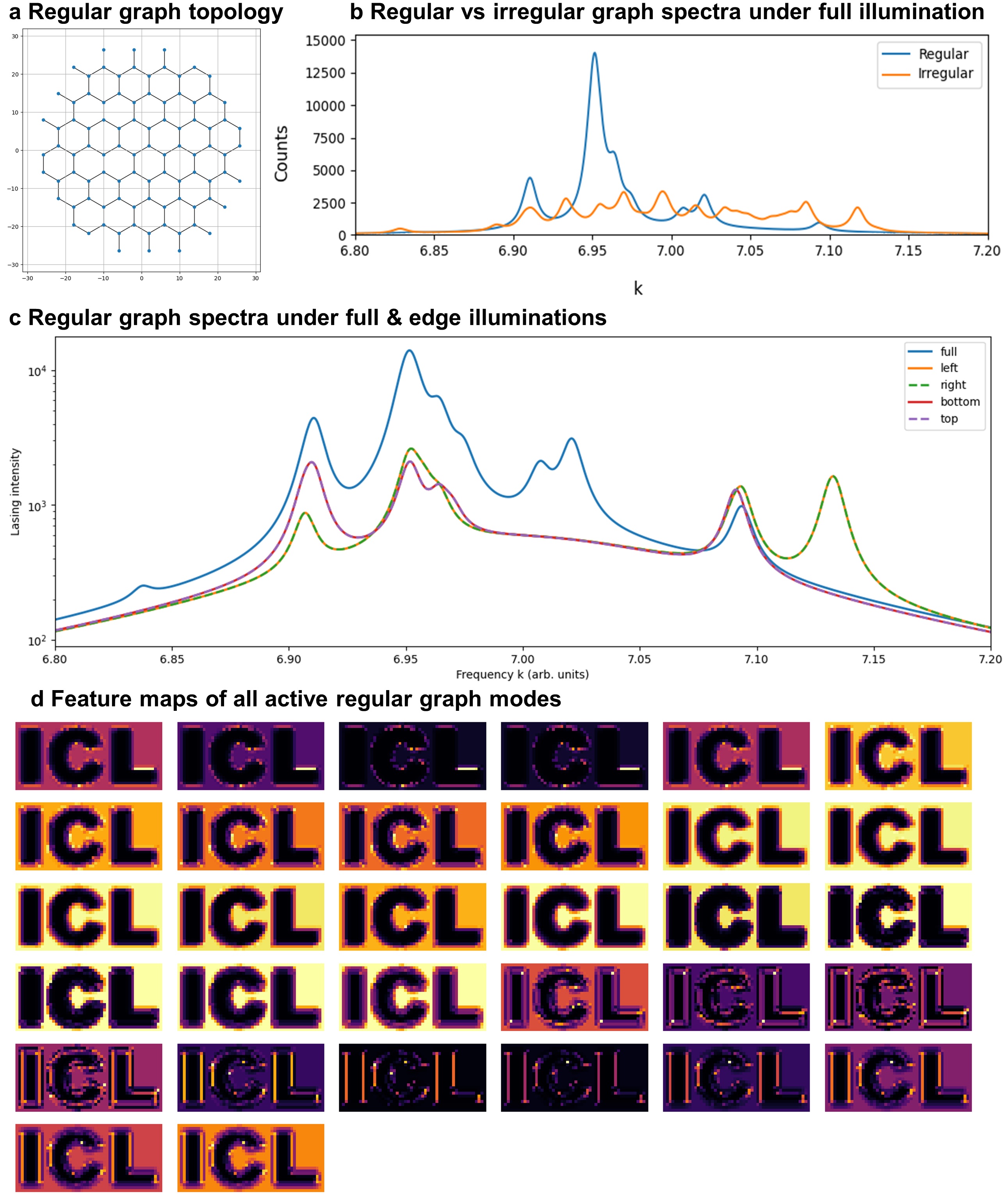}
\caption{netSALT simulations of a regular hexagonal network. \textbf{(a)} Regular hexagonal graph topology, 50 micron diameter graph. \textbf{(b)} Comparison of network spectra under full illumination between the regular hexagonal network and the irregular network examined in fig 2. \textbf{(c)} Comparison of regular hexagonal network lasing spectra under a variety of illumination patterns, full illumination and left, right, bottom and top edge illumination. The spectral response can differentiate between full illumination, left/right edge illumination and top/bottom edge illumination - but not between left and right edges, or top and bottom edges. The irregular network provides more sensitivity across a greater range of image features than the regular network.
\textbf{(d)} Mode feature maps for all the active regular graph modes. Far fewer features are detected than the irregular graph.
}
\label{netSALT-regular} 
\end{figure}

Figure \ref{netSALT-regular} shows netSALT simulated lasing spectra and feature detection for a network with a regular hexagonal graph topology. The regular network shows some ability to detect features, seen in figures \ref{netSALT-regular} c,d) where the system can produce different outputs between full illumination, left/right edge illumination and top/bottom edge illumination - but the regular network can not discern the difference between left and right edges or top and bottom edges. Looking at the full range of detected feature maps in \ref{netSALT-regular}d), a much smaller range of features are detected relative to the irregular network.

\newpage
\subsection*{Experimental feature detection - additional feature maps and test images}
\addcontentsline{toc}{subsection}{Experimental feature detection - additional feature maps and test images}

\begin{figure}[htbp]
\centering
\includegraphics[height=0.9\textheight]{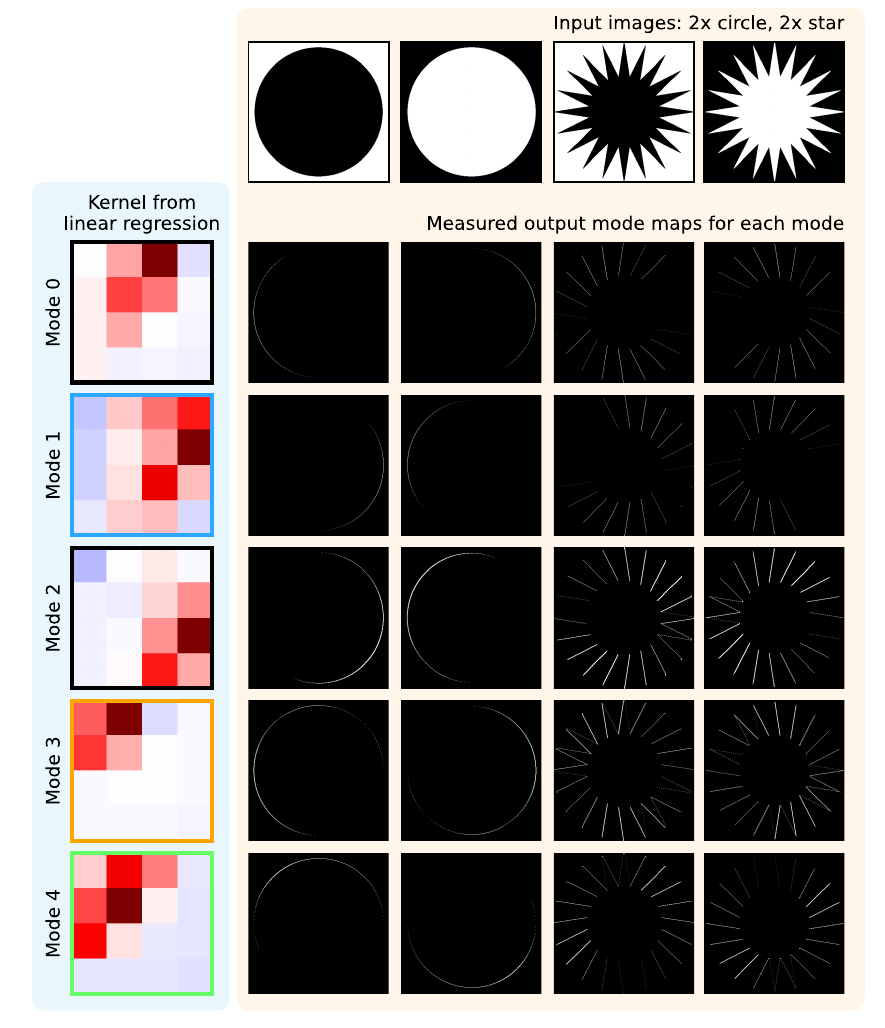}
\caption{Experimental feature maps and learned equivalent kernels for modes 0-5 on four test images, of black and white circles and stars. Each mode has a kernel displayed which was learned by comparing input images to experimentally measured feature maps, and learning the effective kernel via linear regression.}
\label{kernels_starcircle1}
\end{figure}

\begin{figure}[htbp]
\centering
\includegraphics[height=0.9\textheight]{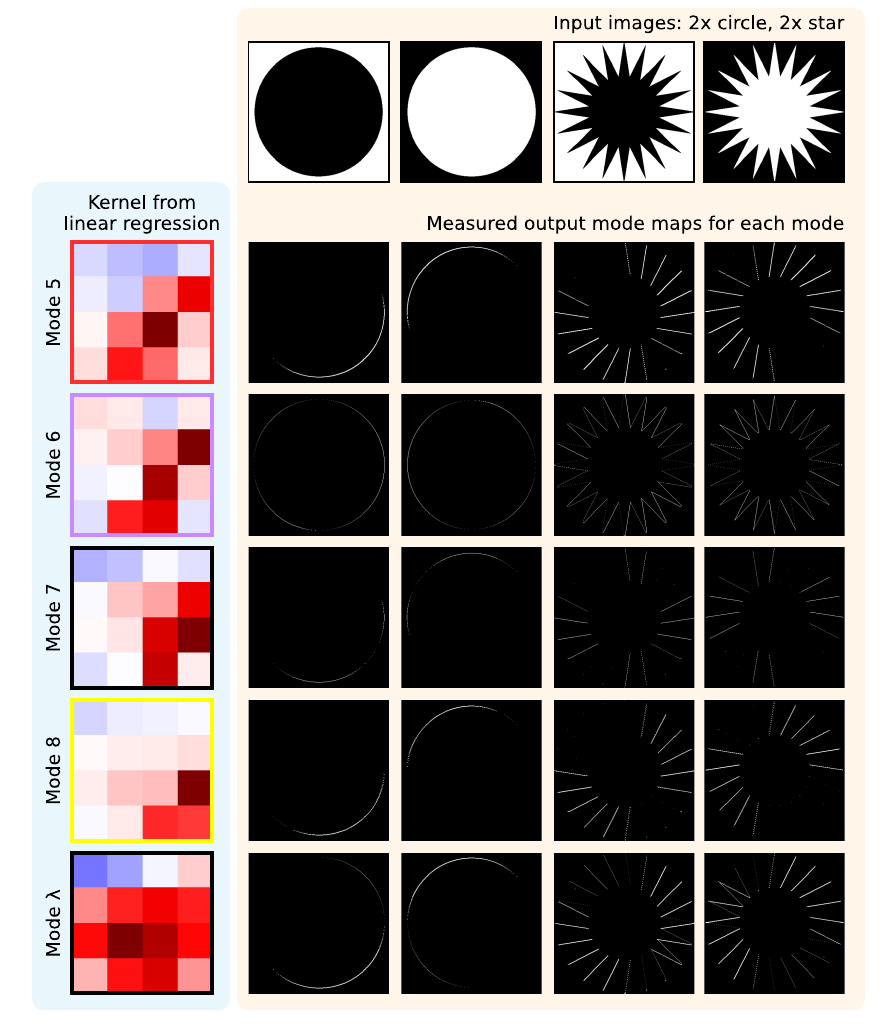}
\caption{Experimental feature maps and learned equivalent kernels for modes 6-9 (mode 9 integrates over all spatial channels, hence is labelled lambda as it contains no spatial information). on four test images, of black and white circles and stars. Each mode has a kernel displayed which was learned by comparing input images to experimentally measured feature maps, and learning the effective kernel via linear regression. }
\label{kernels_starcircle2}
\end{figure}

\begin{figure}[htbp]
\centering
\includegraphics[height=0.95\textheight]{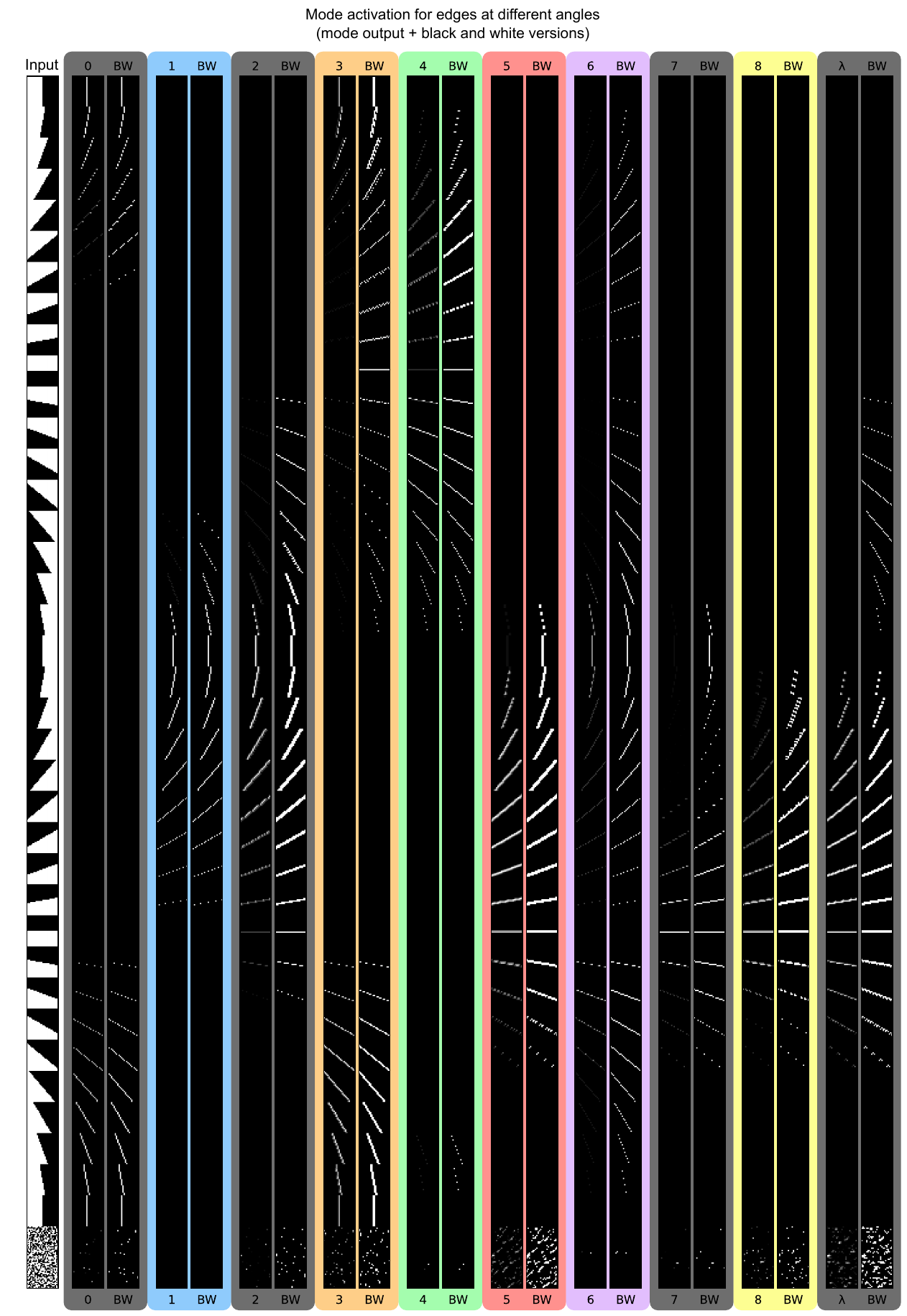}
\caption{Experimental feature maps for modes 0-9 for input images (left-most column) of a range of rotated edges, and two white noise image inputs (bottom two rows). Each mode has two columns, an analogue-varying output produced by the mode-amplitude (left column for each mode, labelled with mode number) and a binarised black and white output produced by normalising the analogue output to 0 or maximum pixel intensity (right column for each mode, labelled BW.)}
\label{kernels_wave}
\end{figure}

Figures \ref{kernels_starcircle1}, \ref{kernels_starcircle2} and \ref{kernels_wave} show experimentally measured feature maps for 10 lasing modes for a range of test images to further examine the feature detection functionality.

Figures \ref{kernels_starcircle1} and \ref{kernels_starcircle2} show feature maps for modes 0-4 and 5-9 respectively in response to four test images: positive and negative tone circles and stars. The feature maps for these test images give a strong visual representation of the range of image features which each mode is sensitive to. Each mode responds to a distribution of similar edges, e.g. some finite arc on the circle or a range of adjacent spikes on the star. This demonstrates that the modes don't only respond to one very specific illumination pattern, but rather a distribution of similar features. This is exactly the behaviour desired for feature detection functionality, eg mode 3 responds strongly across a range of `left' image edges, which is far more useful than only responding to a perfect vertical left edge.

Figure \ref{kernels_wave} shows the response of modes 0-9 for a range of image edges (input images in left-most column) which are gradually rotated. This provides a good visual indication of the range of image features detected by each mode, and also shows that the ten modes provide feature detection response across all angles of image edge between them.

It is important to note that while in netSALT we can spectrally isolate the lasing response of specific modes, even if multiple modes may occupy overlapping spectral regions - this is not possible in experiment. As a result, the selected experimental hyperspectral regions and corresponding feature maps may be contributed to by more than one lasing mode.

\begin{figure}[htbp]
\centering
\includegraphics[height=0.9\textheight]{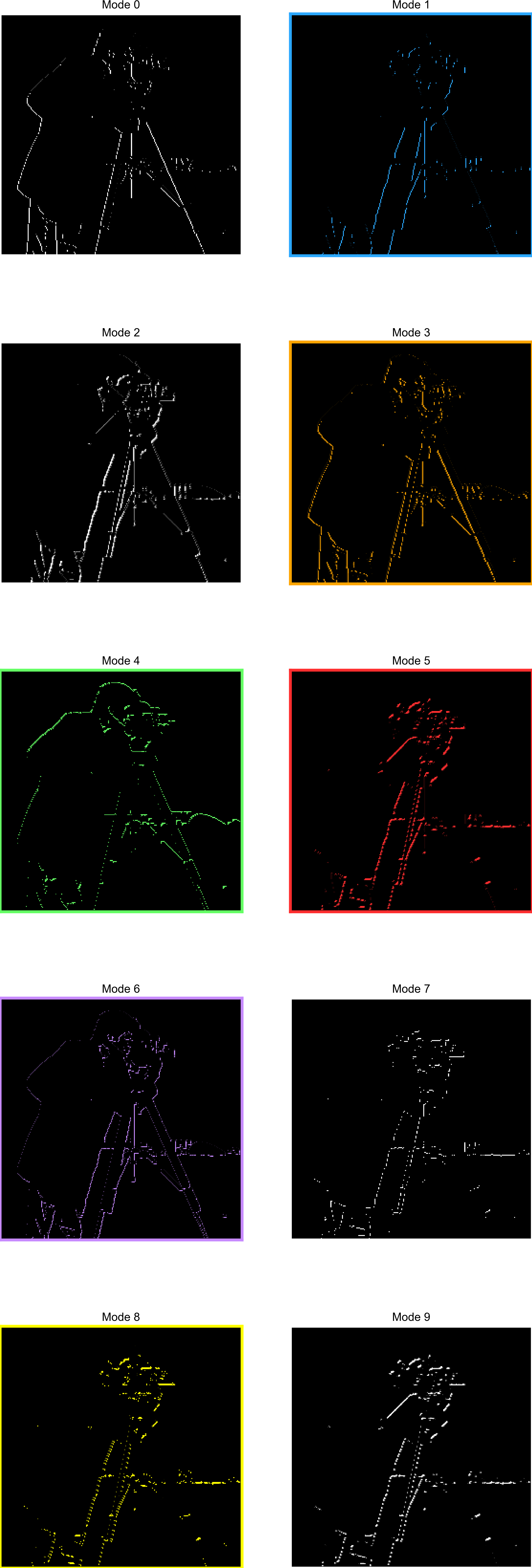}
\caption{Experimental feature maps of the `cameraman' test image for modes 0-9.}
\label{allcameraman}
\end{figure}

\begin{figure}[htbp]
\centering
\includegraphics[height=\textheight]{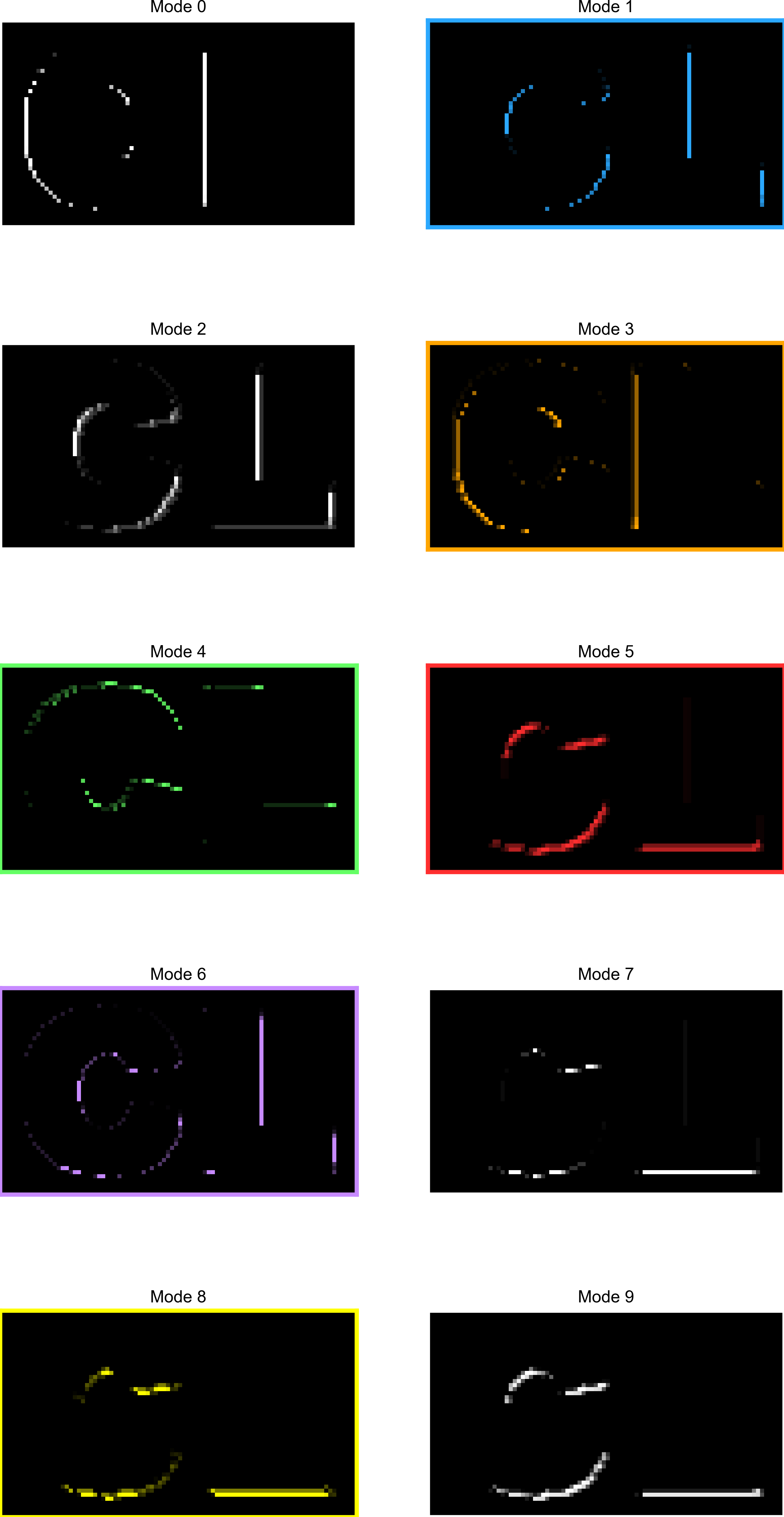}
\caption{Experimental feature maps of `CL' test image for modes 0-9.}
\label{allCL}
\end{figure}

Figures \ref{allcameraman}, \ref{allCL}, \ref{allStreet}, \ref{allDigits} and \ref{allFashion} show feature maps for all ten modes across the cameraman, CL and street test images, and selected MNIST-digits and Fashion-MNIST images respectively.

Figures \ref{feature_maps_rgb_k4}, \ref{feature_maps_rgb_k7}, and \ref{feature_maps_rgb_k11} show the individual mode maps used to construct the composite RGB maps in Fig.~\ref{Fig2sketch}~k.

\begin{figure}[htbp]
\centering
\includegraphics[height=0.35\textwidth]{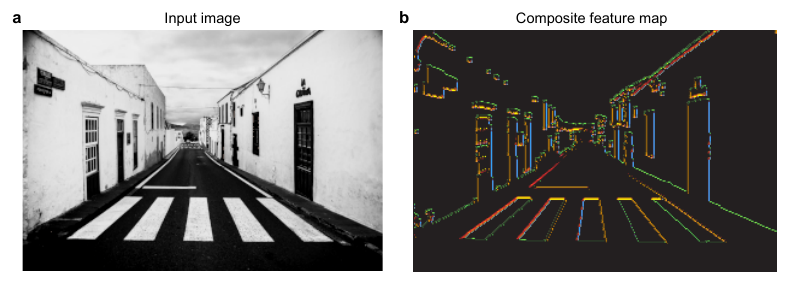}
\caption{`Street' test image (a) and experimental composite feature map (b).}
\label{StreetMain}
\end{figure}

\begin{figure}[htbp]
\centering
\includegraphics[height=0.9\textheight]{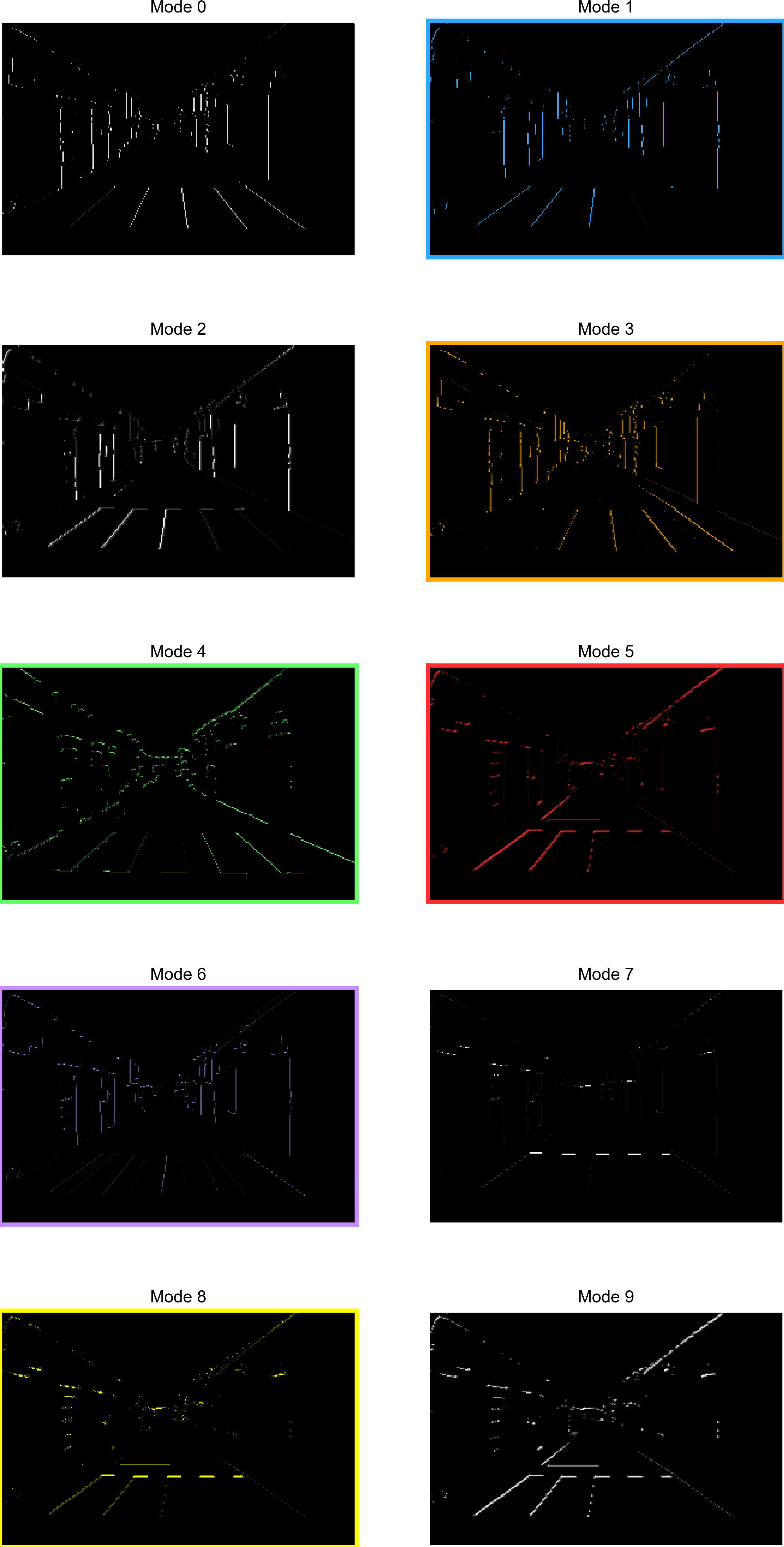}
\caption{Experimental feature maps of `street' test image for modes 0-9.}
\label{allStreet}
\end{figure}

\begin{figure}[htbp]
\centering
\includegraphics[width=\textwidth]{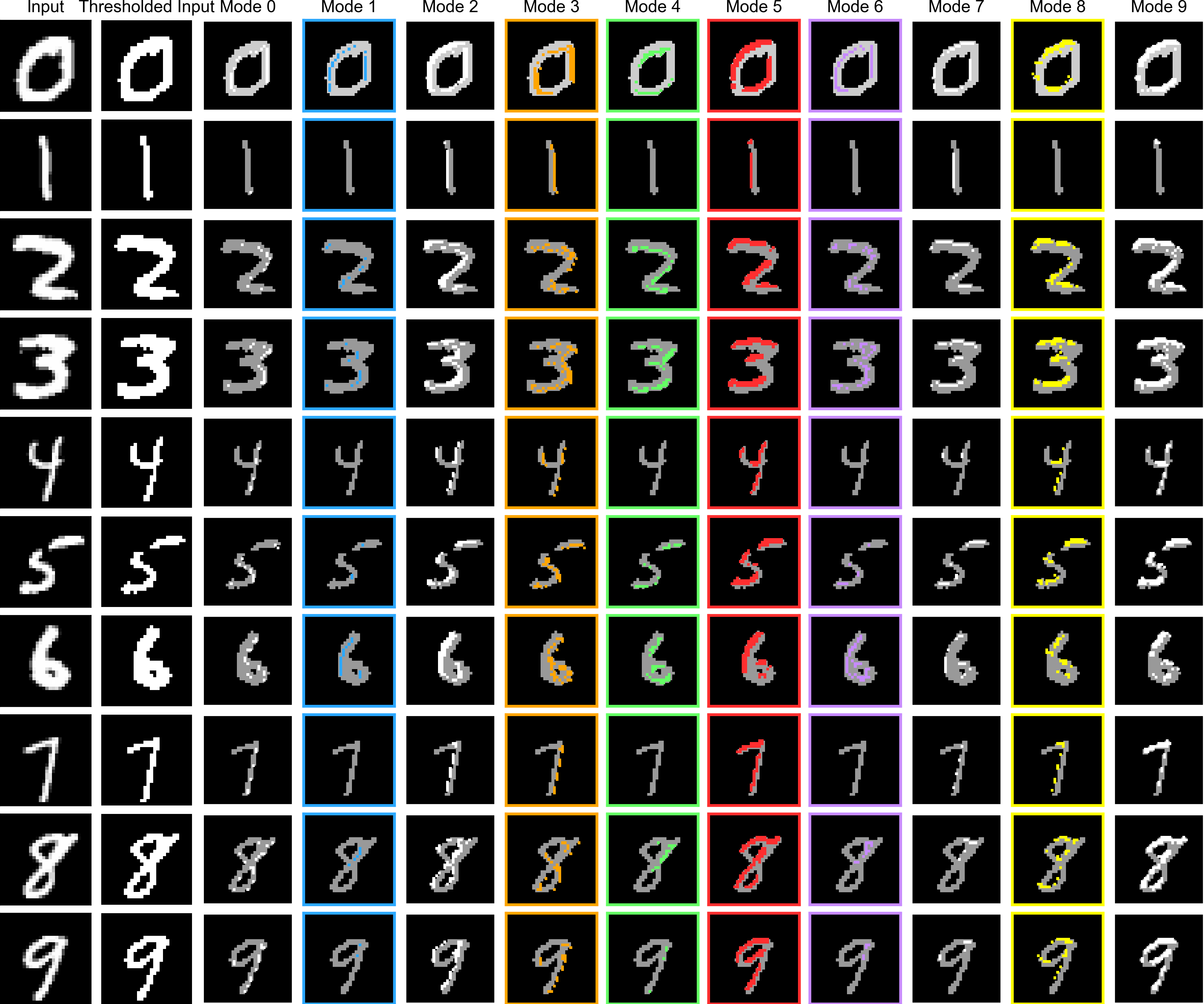}
\caption{Experimental feature maps of ten MNIST digits test images, one per digit class, for modes 0-9.}
\label{allDigits}
\end{figure}

\begin{figure}[htbp]
\centering
\includegraphics[width=\textwidth]{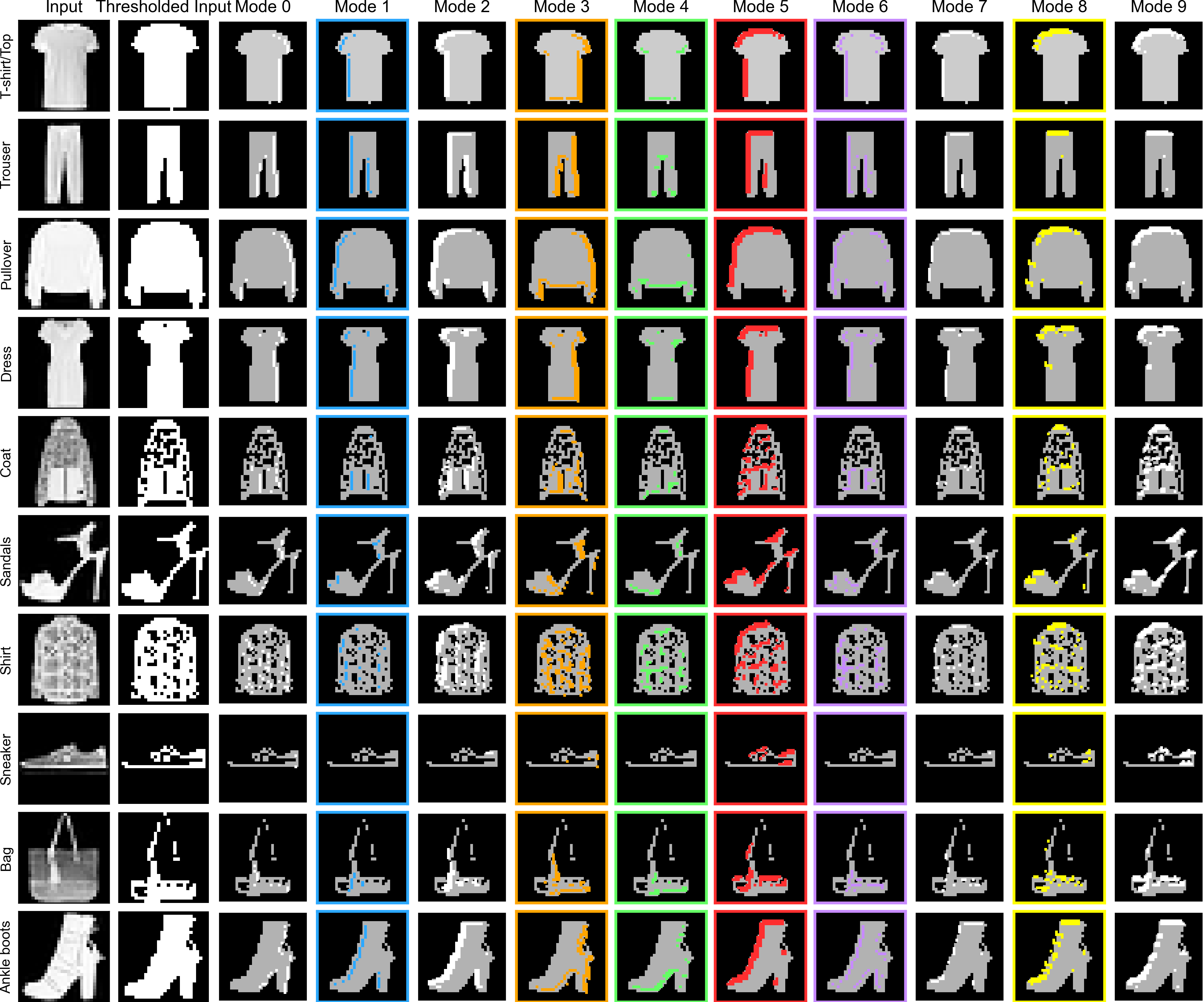}
\caption{Experimental feature maps of ten Fashion-MNIST test images, one per object class, for modes 0-9.}
\label{allFashion}
\end{figure}

\begin{figure}[htbp]
\centering
\includegraphics[width=\textwidth]{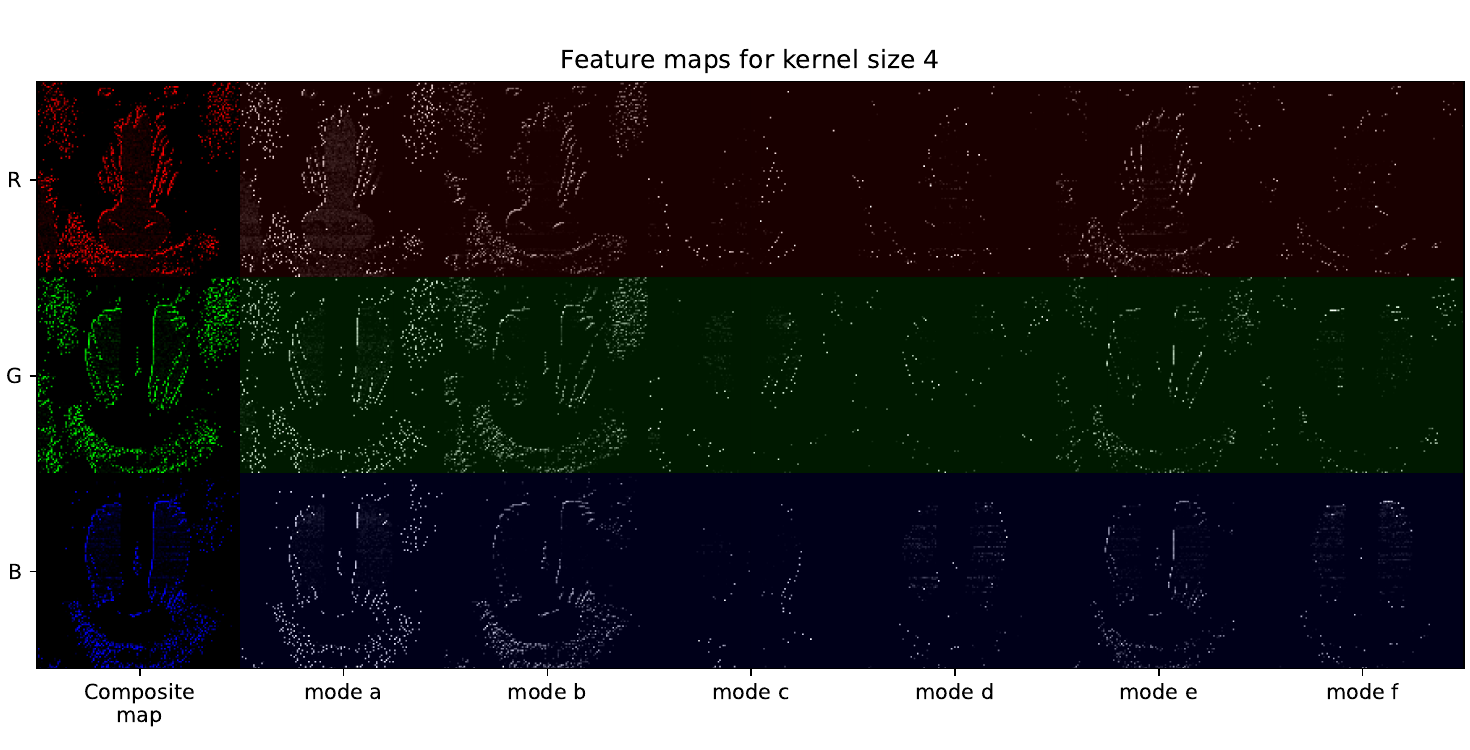}
\caption{Experimental feature maps of for the mandril test image scanned with a 4x4 raster window. Separate scans were performed
for the red, green, and blue image channels, but each scan uses the same modes for the feature detection.}
\label{feature_maps_rgb_k4}
\end{figure}

\begin{figure}[htbp]
\centering
\includegraphics[width=0.8\textwidth]{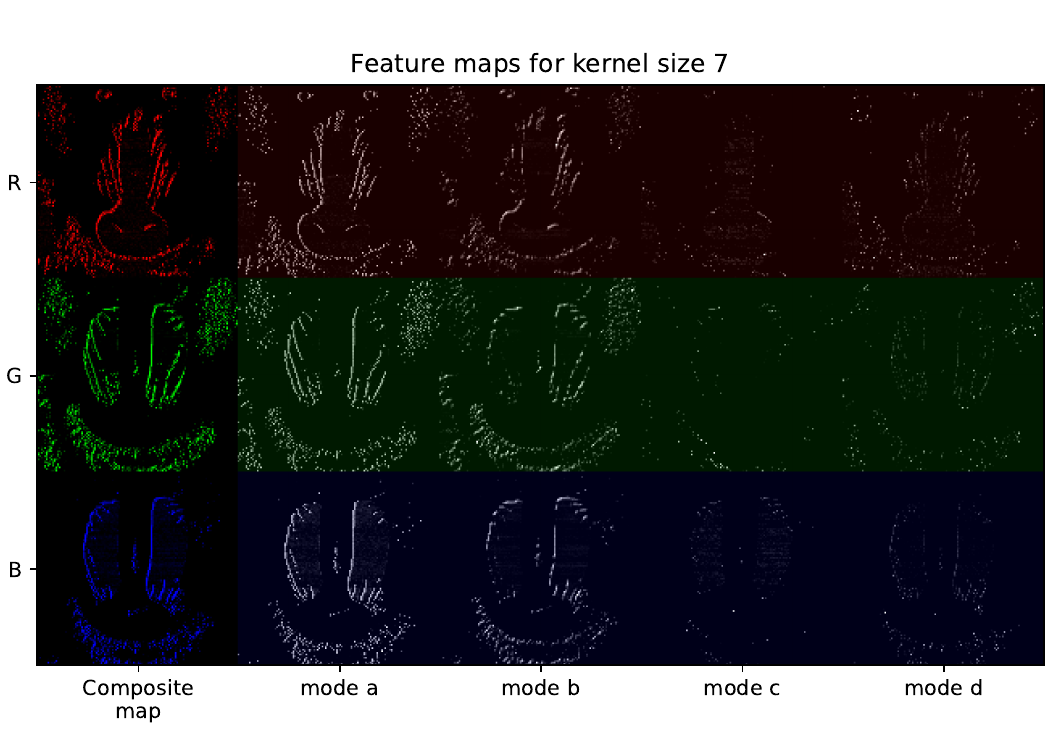}
\caption{Experimental feature maps of for the mandril test image scanned with a 7x7 raster window. Separate scans were performed
for the red, green, and blue image channels, but each scan uses the same modes for the feature detection.}
\label{feature_maps_rgb_k7}
\end{figure}

\begin{figure}[htbp]
\centering
\includegraphics[width=0.8\textwidth]{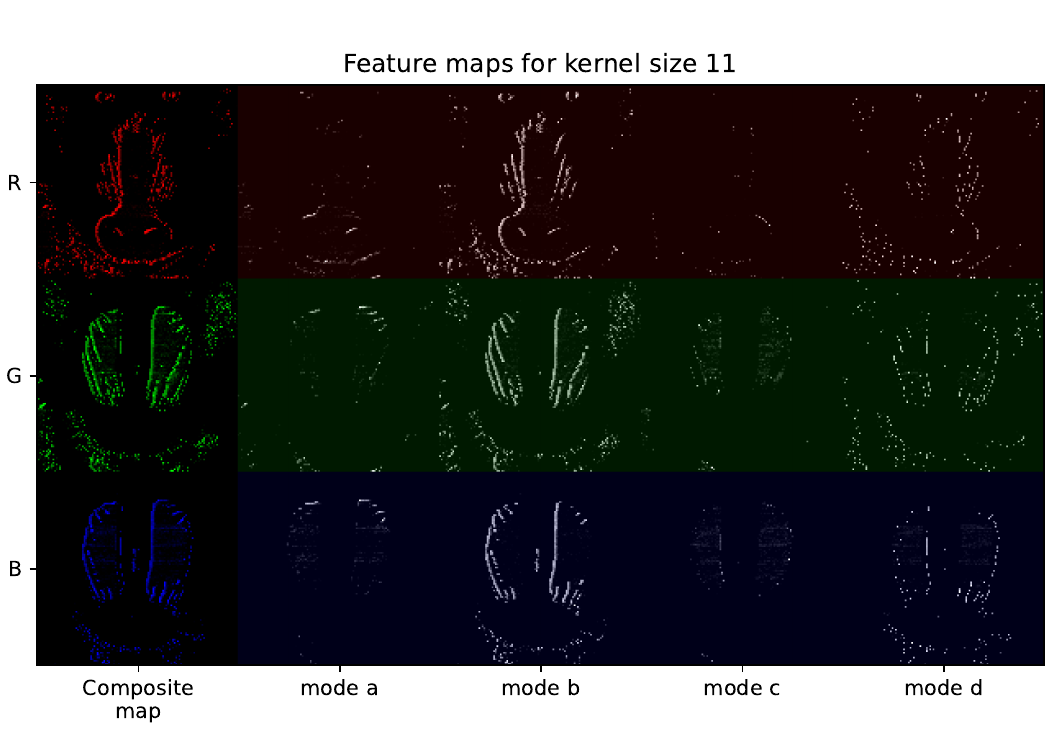}
\caption{Experimental feature maps of for the mandril test image scanned with a 11x11 raster window. Separate scans were performed
for the red, green, and blue image channels, but each scan uses the same modes for the feature detection.}
\label{feature_maps_rgb_k11}
\end{figure}

\clearpage
\subsection*{Least-squares fitting of kernel filters}
\addcontentsline{toc}{subsection}{Least-squares fitting of kernel filters}

\begin{figure}[htbp]
\centering
\includegraphics[width=0.95\textwidth]{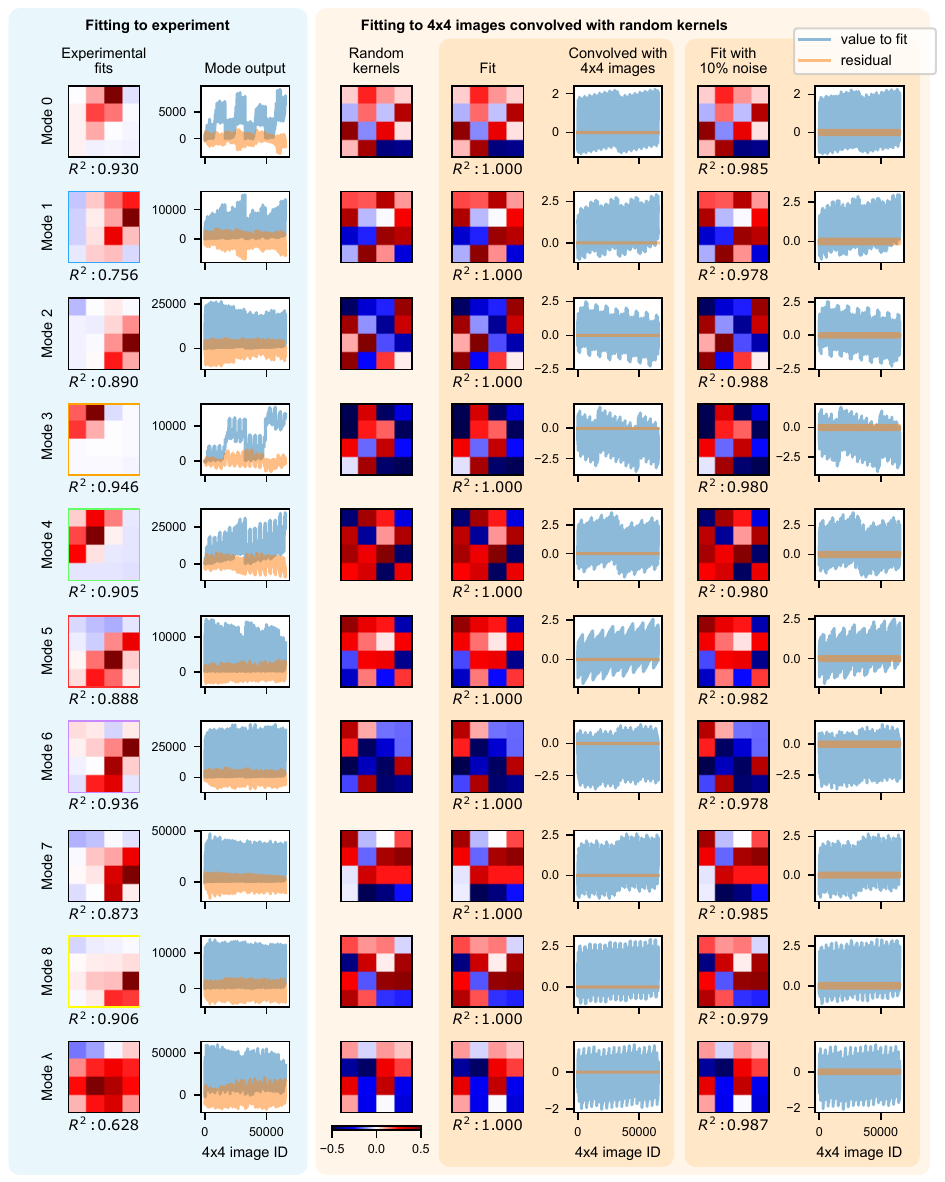}
\caption{Least-squares fitting of kernel filters. Left-hand blue column shows the results of the fitting process for the ten experimental feature detecting modes. R$^2$ values from 0.628-0.946 are found, suggesting aspects of the experimental feature detection functionality are not fully captured by the linear fitting. The orange and blue mode output traces show the target value to be fitted (blue) and the fitting error residual (orange) for the 65,536 ($2^{16}$) possible $4\times4$ pixel images. The central orange column shows the same process performed for randomly generated linear software kernel filters, showing perfect R$^2$ scores of 1.0. The right-hand column shows the same process performed where Gaussian noise of 10\% amplitude relative to the maximum kernel weight is added, with reduced but still high R$^2$ scores of 0.978-0.988.}
\label{fitted_kernels}
\end{figure}

As described in the main text and shown in figure \ref{Fig2sketch}i), we perform linear least-squares fitting between input images and experimentally measured mode amplitudes to fit effective linear kernel filters for the experimental modes performing feature detection. Results of this process are shown in figure \ref{fitted_kernels}. The fitting process uses a least-squares method to find the linear kernel filter weights which best reproduce the filter output (e.g. the lasing mode amplitude for our experimental case) for a given set of input images and filter outputs. As our input image set, we use the 65,536 ($2^{16}$) possible $4\times4$ black and white pixel images.

We plot the results of the fitting process for our ten experimental feature-detecting modes in the blue left-hand column of figure \ref{fitted_kernels}, titled `experimental fits'. For the 65,536 input images we show the target output (eg. the experimental lasing mode amplitude) and residual fitting errors in the blue and orange traces respectively in the column titled `mode output' next to the fitted kernels. Substantial residual errors are observed, with R$^2$ values of 0.628-0.946. This suggests that the linear fitting process may be unable to fully capture the feature detecting dynamics of the experimental system. 

To test this assumption, we also performed the same fitting process on linear software kernel filters, initialising ten $4\times4$ filters with randomly generated weights. The random kernels are plotted in the left-hand orange column titled `random kernels'. The results of the linear fitting process for these kernels are shown in the adjacent columns titled `fit' (the kernel weights) and `convolved with 4$\times$4 images' (target and residual errors). Here, perfect R$^2$ scores of 1.0 are observed, as expected when performing a linear fit on a linear process. To test whether the poorer-quality linear fits to our experimental data are a result of experimental noise, we performed a further test where we added Gaussian noise to the software filter output. We used a noise amplitude of 10\% relative to the maximum software kernel weights. Results of this test are shown in the two right-most columns. Lower $R^2$ of 0.978-0.988 are observed relative to the no-noise case, but still substantially higher than the fits to the experimental data.

\begin{figure}[htbp]
\centering
\includegraphics[width=0.7\textwidth]{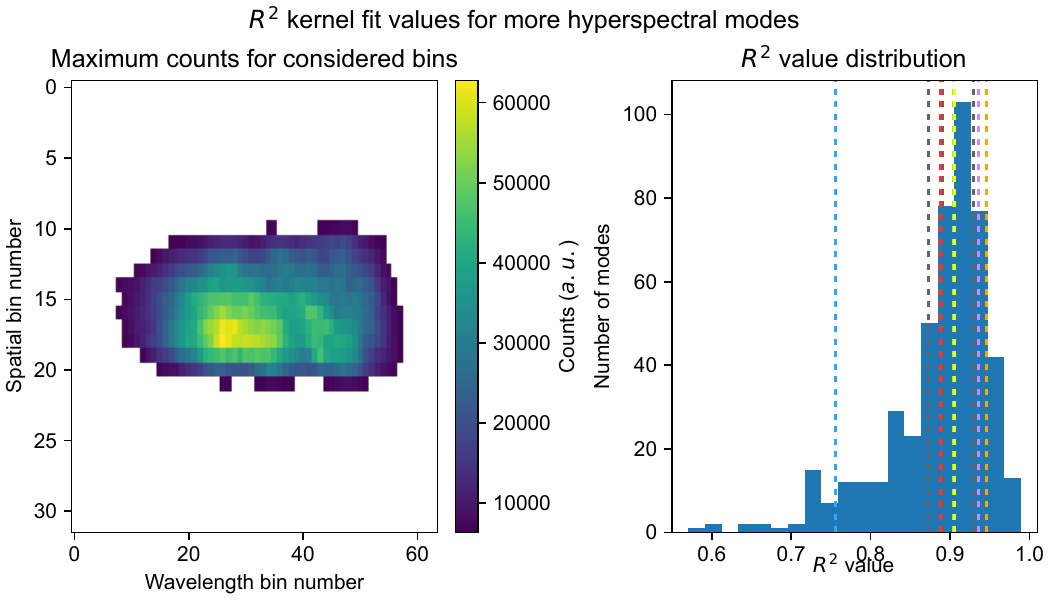}
\caption{Experimental $R^2$ fits for all binned hyperspectral modes that reach more than 10\% of the maximum signal amplitude.
Dashed lines show $R^2$ values for modes in Supplementary Figure~\ref{fitted_kernels}.
}
\label{r2s_all_modes}
\end{figure}

Figure \ref{r2s_all_modes} shows the distribution of $R^2$ values for fits beyond the 10 modes highlighted as feature-detecting in the manuscript. For hyperspectral bins where noise overwhelms the signal, the $R^2$ value would naturally be low, so this analysis is constrained to modes that reach a maximum amplitude of at least 10\% of the overall maximum for the 65,536 input images. The distribution shows that all analysed modes have $R^2$ values lower than 1, with a peak around 0.91 and a long tail towards 0.6. Some nonlinear behaviour is thus exhibited by the majority of modes.

\begin{figure}[htbp]
\centering
\includegraphics[width=0.7\textwidth]{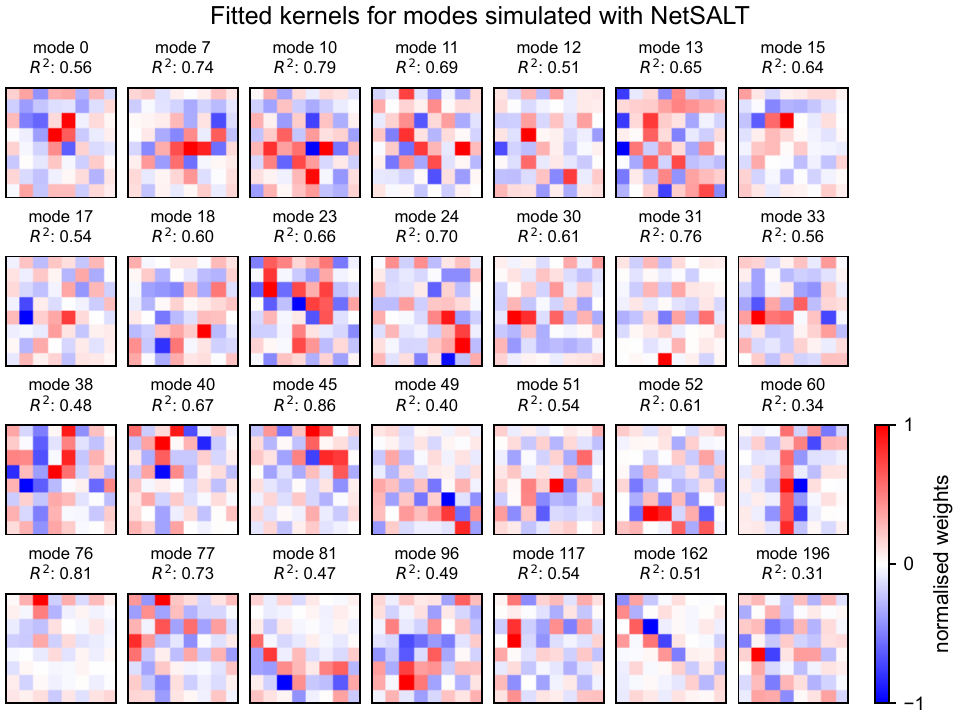}
\caption{Fitted kernels for the netSALT-simulated modes shown in Figure~\ref{Full-netSALT-modes}.
}
\label{netsalt_kernel_fits}
\end{figure}

A similar linear fitting study was performed for the simulated netSALT results, and Figure \ref{netsalt_kernel_fits} shows a set of linearly-fitted kernels for the netSALT modes shown in Figure \ref{Full-netSALT-modes}. The $R^2$ values seen are significantly lower than 1, indicating that the linear fit is unable to fully capture the feature detection process in the simulated network, just as with the experimental one. Negative weights are also observed, illustrating the suppressive effects of mode competition. 401 simulated lasing responses to 8$\times$8 raster slices of the CL image from Figure ~\ref{Fig2sketch}~e were used as the input for fitting, which is smaller than the 65,536 experimental samples, but spans the 64-dimensional input space.

\textbf{}

\clearpage
\subsection*{Quasi-isomorphic software benchmarks}
\addcontentsline{toc}{subsection}{Quasi-isomorphic software benchmarks}

To compare our system against a software network with similar architecture, we constructed a network with quasi-isomorphic structure to our experimental system. The quasi-isomorphic software benchmark comprises an initial convolutional layer with 10 4$\times$4 convolutional filters, followed by a hidden layer of 30,000 neurons with randomised weights and ReLu nonlinear activation, with a final logistic regression layer. We compared our experimental system against two versions of this quasi-isomorphic software network: one with randomised kernel filter weights, and one with kernel filter weights obtained via fits to our experimental data as shown in figure \ref{Fig2sketch}f) and \ref{kernels_starcircle1}, \ref{kernels_starcircle2}. 

\begin{figure}[htbp]
\centering
\includegraphics[width=\textwidth]{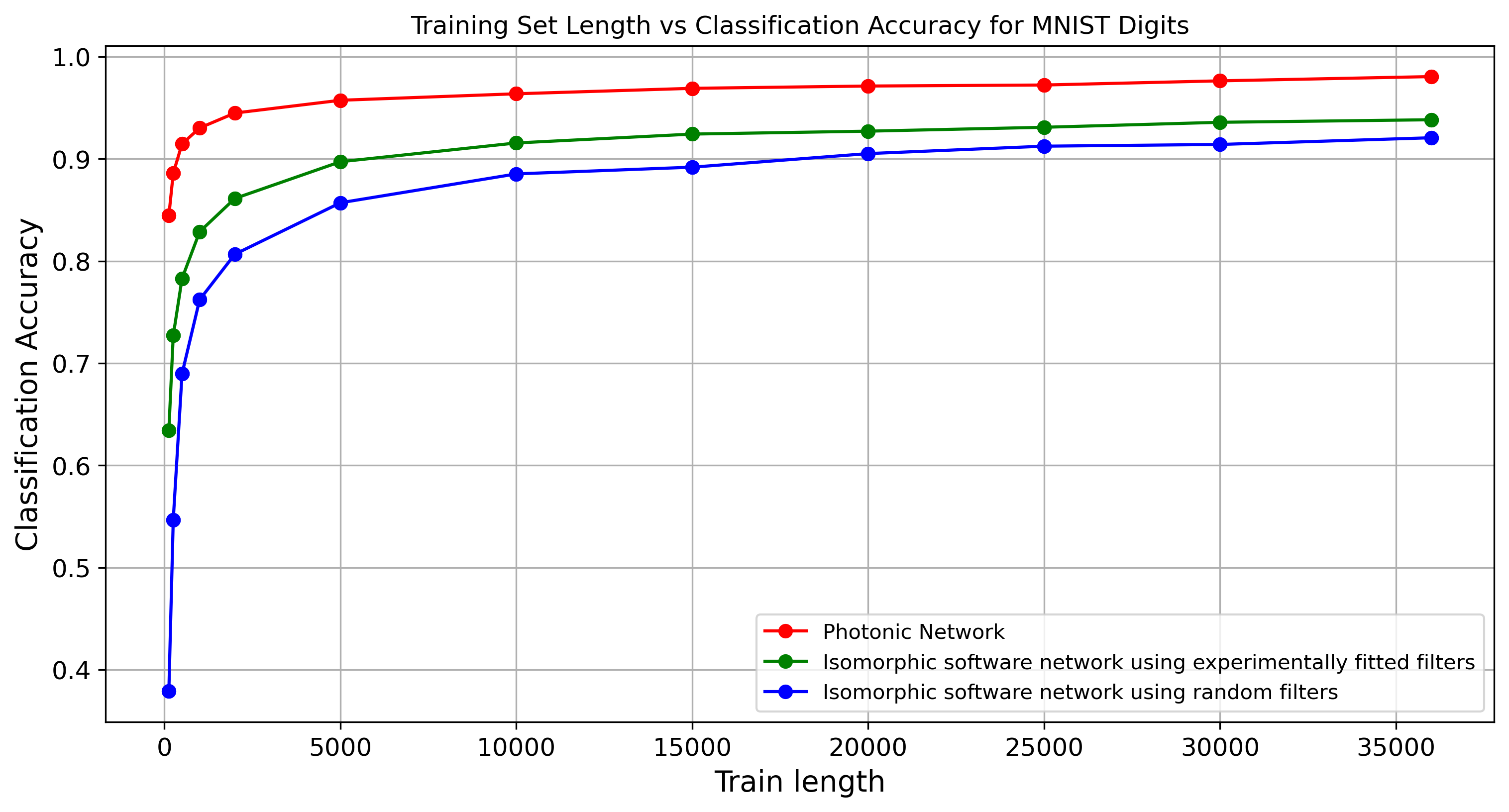}
\caption{MNIST digits classification results for the experimental photonic system and two versions of the quasi-isomorphic software benchmark, one with random kernel filter weights and one with kernel filter weights obtained via fits to our experimental data.}
\label{iso-digits}
\end{figure}

\begin{figure}[htbp]
\centering
\includegraphics[width=\textwidth]{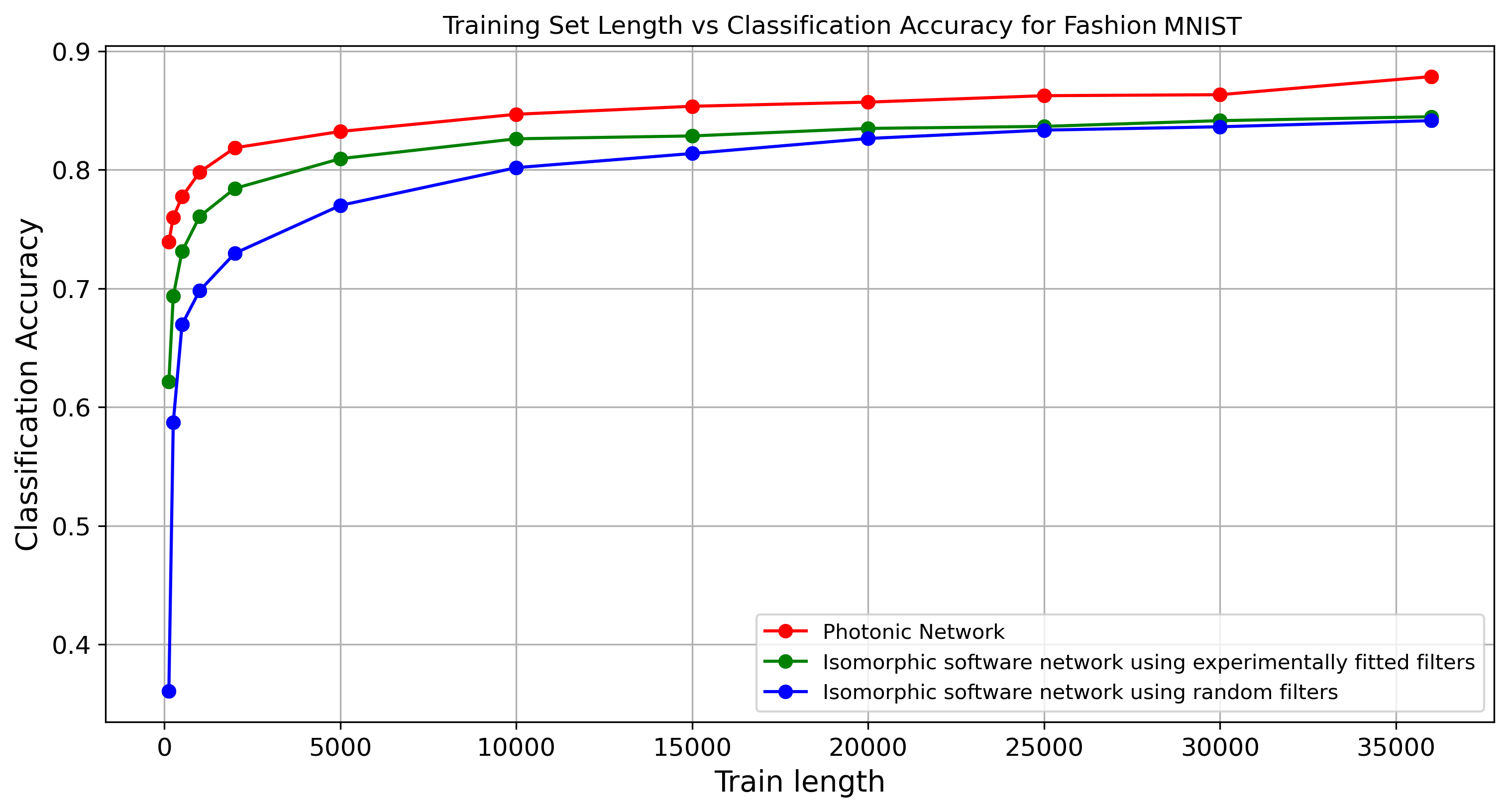}
\caption{Fashion MNIST classification results for the experimental photonic system and two versions of the quasi-isomorphic software benchmark, one with random kernel filter weights and one with kernel filter weights obtained via fits to our experimental data.}
\label{iso-fashion}
\end{figure}

Figures \ref{iso-digits} and \ref{iso-fashion} show the classification results of these tests for the MNIST digits and Fashion-MNIST tasks respectively. The experimental photonic system outperforms the quasi-isomorphic benchmark in both cases, with the software benchmark using the fitted experimental kernel weights performing better.

\subsection*{Speed, footprint, energy and scalability}
\addcontentsline{toc}{subsection}{Speed, footprint, energy and scalability}

\textbf{Speed} The experimental throughput for the results in this paper is up to 100 Hz. While the manuscript was in revision, we confirmed that our system can operate up to 1 kHz using an upgraded camera (Tucsen Aries 6510), recording lasing hyperspectra from the network. This was tested after collecting all the experimental data for this paper, and these faster results are not featured but will be further explored for subsequent works on our system. Beyond this 1 kHz rate, our next limit is our DMD which operates at 10 kHz. Commercial MHz DMDs are available, and DMD technology is currently rapidly advancing. For detection, commercially available Si or InGaAs photodiode arrays (e.g. Hamamatsu) run from $10^4-10^6$ Hz and are readily integrable with our scheme. Beyond this, the photonic dynamics of the system operate at 10-100 ps - going faster than this would require substantial re-engineering of the network material and system architecture, but these 10-100 GHz speeds are considerable. For reference, a modern NVIDIA 4090 GPU running a large CNN, ResNet or EfficientNet equivalent, takes about 1 ms per inference - e.g. 1 kHz operation, and consumes 600 W without considering cooling, RAM, memory buffering etc overheads.

\textbf{Footprint} Our current experimental optical system occupies an area of approximately 450 mm $\times$ 450 mm, with an external fs laser of size of 730 mm $\times$ 419 mm. The fs laser is guided into the optical system to pump the 150 $\upmu$m diameter on-chip InP network laser, with the network providing all photonic nonlinear processing which drives computation. This is similar to other on-chip photonic computing schemes, such as large area vertical cavity surface emitting laser (LA-VCSEL) ($\sim$20 um in diameter)\cite{porteCompleteParallelAutonomous2021} , with substantially reduced on-chip footprint relative to cross-bar arrays ($>$7 mm $\times$ 8 mm)\cite{feldmann2021parallel} , and Mach–Zehnder interferometers ($\sim$6 mm $\times$ 5.5 mm) \cite{bandyopadhyaySinglechipPhotonicDeep2024}  -- which also currently require external laser sources. It is worth noting that the cross-bar and Mach-Zehnder schemes provide degrees of on-chip programming. In future, on-chip programming (electro-optical, etc.) may be introduced to the network laser scheme.

To miniaturise the system in future, a more compact optical path design can be implemented to reach an off-chip optics footprint of $<$100 cm$^{2}$. As discussed in the Methods, the system can also be operated with a 80 ps diode pumped MOPA laser (180 mm $\times$ 90 mm), which is approximately 1/18th of the size of the fs laser. With further steps, future iterations could replace the off-chip pump laser with pixelated on-chip light sources, e.g., VCSEL and microLED arrays, to further reduce the footprint.

\textbf{Energy} Our experimental results for this paper were collected using a system with optical and electronic devices: A 200 fs laser source with an average power of 43 $\upmu$W (pulse energy of 43 nJ) on the network (wall-plug power of 600 W), a DMD and a camera with operating power of 2.7 W and 5 W, respectively. Considering the 100 Hz operational speed of the system, the effective optical energy consumption is 0.43 $\upmu$J/frame, with the effective energy consumption of the whole system summed up to $\sim$77 mJ/frame.
Since the digital regression layer at the end of our inference pipeline is lightweight, amounting to one matrix multiplication per illumination per class, we estimate the power used by the logistic regression step to be on the order of $\sim$0.5 $\upmu$J (single layer) - 5 $\upmu$J (two layer) per frame.

To put the pump energy in perspective, other photonic computing methods (with similar electronic system setup and off-chip power consumption), e.g., multimode fibre takes 8.7 $\upmu$J/frame \cite{oguzProgrammingNonlinearPropagation2024}, LA-VCSEL takes 27.7 $\upmu$J/frame\cite{porteCompleteParallelAutonomous2021}, and diffractive media takes 12.5 mJ/frame\cite{wangLargescalePhotonicComputing2024}.

Improving the efficiency of the pump laser will allow us to greatly improve power and energy consumption. To this end, we have confirmed during revision of the manuscript that a much lower power (10.5 W wall-plug power) 80 picosecond pulse pump laser (Picophotonics CP32, 532 nm) can drive our network to lase, with experimental confirmation shown in supplementary figure \ref{picophotonic-pulse}. Using this pump laser brings our wall-plug power consumption to 16.2 W, which bodes well for future iterations of our scheme using lower power laser and faster kHz throughput.

To put these numbers in perspective, an NVIDIA 4090 GPU consumes 600 W. For the BreaKHis biomedical task considered in this paper, a 4090 GPU running EfficientNetV2-B0 required 16 hours of runtime with 20$\times$ data augmentation (e.g. artificially making 20 augmented/distorted/rotated copies of each training image) in order to match the accuracy performance of our system, while our system took 4 minutes, including optical experiments and logistic regression training. This is of course a limited use-case, relatively hard vision tasks with small training sets, and for larger dataset machine learning tasks with many 10s-100s of thousands of training examples our system underperforms relative to conventional large modern software neural networks. However, these relatively hard, data-restricted tasks are highly relevant for the edge-AI use cases where neuromorphic computing may have a lot to offer. We envision that in future iterations of our scheme, this range of tasks where we offer competitive performance will be substantially expanded by refinements to the physical system, optical experimental scheme, and training approach.

Beyond using free space pulsed lasers, the system should in future be compatible with on-chip pixelated light sources e.g. micro LED arrays and vertical cavity surface emitting lasers (VCSELs) which have far higher energy-efficiency than free-space pulsed lasers, up to 40-50\% efficiency. This would have the added benefit of removing the requirement for a DMD and greatly reducing the system footprint.

\begin{figure}[htbp]
\centering
\includegraphics[width=\textwidth]{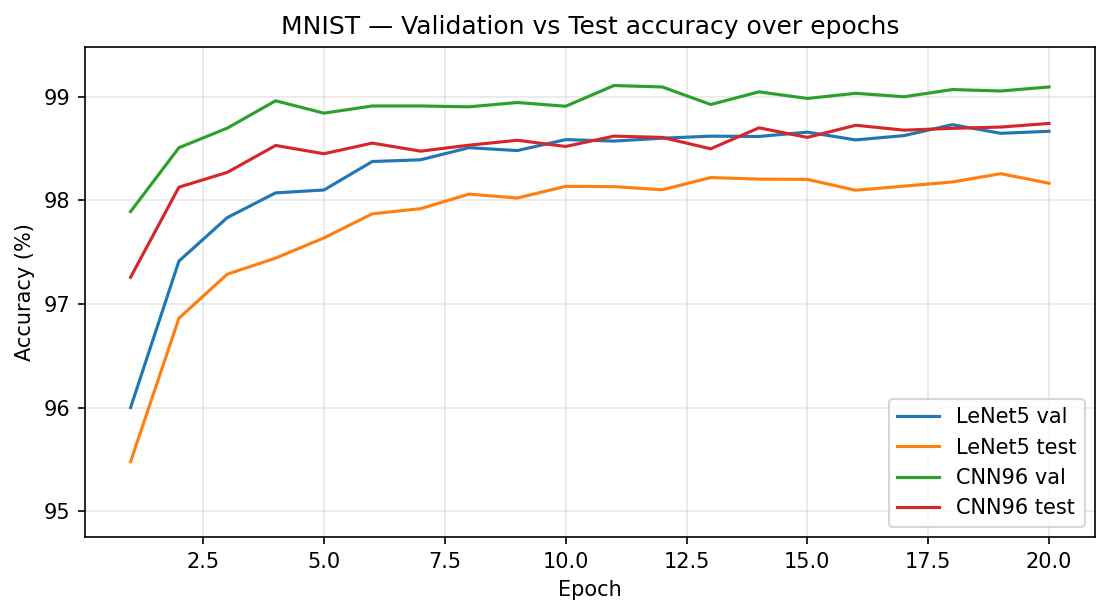}
\caption{Validation vs. test accuracy on MNIST for the `LeNet5' and `96-Filter CNN' software CNN baselines over training epochs. The models converge well with stable plateaus observed.}
\label{MNIST_training_epochs}
\end{figure}

\end{document}